\documentclass[nofootinbib,aps,prl,longbibliography,twocolumn,superscriptaddress,amsmath,amssymb,verbatim,floatfix]{revtex4-2}
\usepackage[utf8]{inputenc}
\usepackage[english,french]{babel}

\usepackage{bm}
\usepackage{xspace}
\xspaceaddexceptions{]} %
\usepackage{braket} 

\usepackage{siunitx} 
\usepackage[T1]{fontenc}
\DeclareSIUnit[number-unit-product = {}]{\inchQ}{\textquotedbl} 
\DeclareSIUnit[number-unit-product = {\thinspace}]{\inch}{in} 

\usepackage{chemformula} 
\usepackage[super]{nth} 
\usepackage{graphicx}
\usepackage{lmodern}
\usepackage{epstopdf}
\AtBeginDocument{\usepackage{booktabs}} 
\usepackage[colorlinks=true,citecolor=blue,linkcolor=blue,urlcolor=blue,bookmarks=true]{hyperref}
\usepackage{xcolor}

\usepackage[nameinlink,capitalize]{cleveref}
\crefname{section}{Sec.}{Secs.}
\crefname{pluralfigure}{Figs.}{Figs.}
\crefname{pluralequation}{Eqs.}{Eqs.}
\creflabelformat{pluralequation}{#2(#1)#3}

\newcommand{\eg}{e.g.\xspace}
\newcommand{\via}{via\xspace}

\newcommand{\abinitio}{\textit{ab initio}\xspace}

\newcommand{\nontrivial}{non-trivial\xspace}

\newcommand{\noncoplanar}{non-coplanar\xspace}

\newcommand{\centrosymmetric}{centrosymmetric\xspace}
\newcommand{\Centrosymmetric}{Centrosymmetric\xspace}
\newcommand{\noncentrosymmetric}{noncentrosymmetric\xspace}

\newcommand{\incommensurate}{incommensurate\xspace}

\newcommand{\dzmLong}{Dzyaloshinskii--Moriya\xspace}

\newcommand{\dmi}{DMI\xspace}
\newcommand{\dmiLong}{\dzmLong interactions\xspace}
\newcommand{\dmiLongFirst}{\dmiLong (\dmi)\xspace}
\newcommand{\rkky}{RKKY\xspace}
\newcommand{\rkkyLong}{Ruderman--Kittel--Kasuya--Yosida\xspace}
\newcommand{\rkkyLongFirst}{\rkkyLong (\rkky)\xspace}

\newcommand{\quasistatic}{quasi-static\xspace}
\newcommand{\Quasistatic}{Quasi-static\xspace}

\newcommand{\wavevector}{wave vector\xspace} 

\newcommand{\wavelength}{wavelength\xspace} 

\newcommand{\bragg}{Bragg\xspace}

\newcommand{\dft}{DFT\xspace}
\newcommand{\dftLong}{density functional theory\xspace}

\newcommand{\dftLongFirst}{\dftLong (\dft)\xspace}

\newcommand{\zfc}{ZFC\xspace}

\newcommand{\zfcIngLong}{zero field cooling\xspace}

\newcommand{\zfcIngLongFirst}{\zfcIngLong (\zfc)\xspace}

\newcommand{\dosLong}{density of states\xspace}

\newcommand{\icOne}{IC-1\xspace}

\newcommand{\icTwo}{IC-2\xspace}

\newcommand{\skl}{SkL\xspace}

\newcommand{\sklLongAdj}{skyrmion-lattice\xspace}

\newcommand{\sklLongAdjFirst}{\sklLongAdj (\skl)\xspace}

\newcommand{\mamLong}{meron--antimeron\xspace}

\newcommand{\mamlLong}{\mamLong lattice\xspace}

\newcommand{\mamlAlt}{ML\xspace} 
\newcommand{\mamlAltLongFirst}{\mamlLong (\mamlAlt)\xspace}

\newcommand{\paramag}{PM\xspace}
\newcommand{\paramagLong}{paramagnetic\xspace}

\newcommand{\paramagLongFirst}{\paramagLong (\paramag)\xspace}

\newcommand{\xray}{X-ray\xspace}

\newcommand{\rxs}{RXS\xspace}
\newcommand{\rxsLong}{resonant \xray scattering\xspace}

\newcommand{\rxsLongFirst}{\rxsLong (\rxs)\xspace}

\newcommand{\arpes}{ARPES\xspace}
\newcommand{\arpesLong}{angle-resolved photoemission spectroscopy\xspace}

\newcommand{\arpesLongFirst}{\arpesLong (\arpes)\xspace}

\newcommand{\muon}{$\mu$\xspace}

\newcommand{\musr}{{\muon}SR\xspace} 

\newcommand{\musrLongSpec}{muon spectroscopy\xspace}

\newcommand{\musrLongSpecFirst}{\musrLongSpec (\musr)\xspace}

\newcommand{\lf}{LF\xspace}

\newcommand{\lfLongAdj}{longitudinal-field\xspace}

\newcommand{\lfLongAdjFirst}{\lfLongAdj (\lf)\xspace}

\newcommand{\tf}{TF\xspace}

\newcommand{\tfLongAdj}{transverse-field\xspace}

\newcommand{\tfLongAdjFirst}{\tfLongAdj (\tf)\xspace}

\newcommand{\zf}{ZF\xspace}

\newcommand{\zfLongAdj}{zero-field\xspace}

\newcommand{\zfLongAdjFirst}{\zfLongAdj (\zf)\xspace}

\newcommand{\tDirEq}{{\mathrm{T}}} 
\newcommand{\tDir}{$\tDirEq$\xspace}
\newcommand{\lambdaTDirEq}{\lambda_\tDirEq} 

\newcommand{\arTDirEq}{a_{r,\tDirEq}} 
\newcommand{\arTDir}{$\arTDirEq$\xspace}

\newcommand{\lDirEq}{{\mathrm{L}}} 
\newcommand{\lDir}{$\lDirEq$\xspace}
\newcommand{\lambdaLDirEq}{\lambda_\lDirEq} 
\newcommand{\lambdaLDir}{$\lambdaLDirEq$\xspace}
\newcommand{\arLDirEq}{a_{r,\lDirEq}} 
\newcommand{\arLDir}{$\arLDirEq$\xspace}

\newcommand{\instrHiFi}{HiFi\xspace}

\newcommand{\instrGPS}{GPS\xspace}

\newcommand{\gdpdsi}{\ch{Gd2PdSi3}\xspace}
\newcommand{\gdAtom}{\ch{Gd}\xspace}
\newcommand{\gdIon}{\ch{Gd^{3+}}\xspace}

\newcommand{\rpdsi}{\ch{$R$2PdSi3}\xspace}
\newcommand{\gdrual}{\ch{Gd3Ru4Al12}\xspace}
\newcommand{\gdrusi}{\ch{GdRu2Si2}\xspace}
\newcommand{\gdruge}{\ch{GdRu2Ge2}\xspace}
\newcommand{\eual}{\ch{EuAl4}\xspace}
\newcommand{\euAtom}{\ch{Eu}\xspace}

\newcommand{\coznmn}{\ch{Co8Zn9Mn3}\xspace}

\newcommand{\threefold}{threefold\xspace} 

\newcommand{\oneDim}{1D\xspace} 
\newcommand{\threeDim}{3D\xspace} 

\newcommand{\approxproptoinner}[2]{%
  \mathrel{%
    \setbox0=\hbox{$#1\sim$}%
    \setbox2=\hbox{%
      \rlap{\hbox{$#1\propto$}}%
      \lower1.1\ht0\box0%
    }%
    \raise0.25\ht2\box2%
  }%
}

\newcommand{\const}{\mathrm{const.}} 

\newcommand{\myvec}[1]{{\boldsymbol{\mathbf{#1}}}} 

\definecolor{darkgreen}{RGB}{0,128,0}
\definecolor{darkblue}{RGB}{0,0,128}
\definecolor{nblue2}{RGB}{24,118,178}
\definecolor{nyellow}{RGB}{205,116,0} 

\usepackage[normalem]{ulem} 

\newcommand{\supp}{SM\xspace}                        

\newcommand{\suppCite}{\supp~\cite{supp}\xspace}

\begin{document}

\selectlanguage{english}

\preprint{APS/123-QED}

\title{Anisotropic Skyrmion and Multi-$q$ Spin Dynamics in \Centrosymmetric \gdpdsi}



\author{M. Gomil\v{s}ek} 
\email{matjaz.gomilsek@ijs.si} 
\affiliation{Jo\v{z}ef Stefan Institute, Jamova c.~39, SI-1000 Ljubljana, Slovenia}
\affiliation{Faculty of Mathematics and Physics, University of Ljubljana, Jadranska u.~19, SI-1000 Ljubljana, Slovenia}
\affiliation{Department of Physics, Durham University, South Road, Durham DH1 3LE, United Kingdom}


\author{T. J. Hicken} 
\affiliation{Laboratory for Muon Spin Spectroscopy, Paul Scherrer Institut, 5232 Villigen PSI, Switzerland}
\affiliation{Department of Physics, Durham University, South Road, Durham DH1 3LE, United Kingdom}

\author{M. N. Wilson} 
\affiliation{Department of Physics and Physical Oceanography, Memorial University, A1B 3X7, Canada}
\affiliation{Department of Physics, Durham University, South Road, Durham DH1 3LE, United Kingdom}


\author{K. J. A. Franke}
\affiliation{School of Physics and Astronomy, University of Leeds, LS2 9JT, United Kingdom}
\affiliation{Department of Physics, Durham University, South Road, Durham DH1 3LE, United Kingdom}


\author{B. M. Huddart} 
\affiliation{Clarendon Laboratory, University of Oxford, Department of Physics, Oxford OX1 3PU, United Kingdom}
\affiliation{Department of Physics, Durham University, South Road, Durham DH1 3LE, United Kingdom}


\author{A. \v{S}tefan\v{c}i\v{c}} 
\affiliation{University of Warwick, Department of Physics, Coventry CV4 7AL, United Kingdom}

\author{S. J. R. Holt} 
\altaffiliation[Current addresses: ]{Faculty of Engineering and Physical Sciences, University of Southampton, Southampton SO17 1BJ, United Kingdom}
\altaffiliation{Max Planck Institute for the Structure and Dynamics of Matter, Luruper Chaussee~149, 22761 Hamburg, Germany}
\affiliation{University of Warwick, Department of Physics, Coventry CV4 7AL, United Kingdom}

\author{G. Balakrishnan} 
\affiliation{University of Warwick, Department of Physics, Coventry CV4 7AL, United Kingdom}


\author{D. A. Mayoh} 
\affiliation{University of Warwick, Department of Physics, Coventry CV4 7AL, United Kingdom}


\author{M. T. Birch} 
\affiliation{Max Planck Institute for Intelligent Systems, Heisenbergstrasse~3, D-70569 Stuttgart, Germany}
\affiliation{RIKEN Center for Emergent Matter Science, JP-351-0198 Wako, Japan}
\affiliation{Department of Physics, Durham University, South Road, Durham DH1 3LE, United Kingdom}

\author{S. H. Moody} 
\affiliation{Laboratory for Neutron Scattering and Imaging, Paul Scherrer Institut, 5232 Villigen PSI, Switzerland}
\affiliation{Department of Physics, Durham University, South Road, Durham DH1 3LE, United Kingdom}


\author{H. Luetkens} 
\affiliation{Laboratory for Muon Spin Spectroscopy, Paul Scherrer Institut, 5232 Villigen PSI, Switzerland}

\author{Z. Guguchia} 
\affiliation{Laboratory for Muon Spin Spectroscopy, Paul Scherrer Institut, 5232 Villigen PSI, Switzerland}

\author{M. T. F. Telling} 
\affiliation{ISIS Facility, STFC Rutherford Appleton Laboratory, Didcot, Oxfordshire OX11 0QX, United Kingdom}

\author{P. J. Baker} 
\affiliation{ISIS Facility, STFC Rutherford Appleton Laboratory, Didcot, Oxfordshire OX11 0QX, United Kingdom}


\author{S. J. Clark} 
\affiliation{Department of Physics, Durham University, South Road, Durham DH1 3LE, United Kingdom}

\author{T. Lancaster} 
\affiliation{Department of Physics, Durham University, South Road, Durham DH1 3LE, United Kingdom}


\date{\today}

\begin{abstract}
Skyrmions are particle-like vortices of magnetization with \nontrivial topology, which are usually stabilized by \dmiLongFirst in \noncentrosymmetric bulk materials. 
Exceptions are \centrosymmetric \gdAtom- and \euAtom-based \sklLongAdjFirst hosts with zero \dmi, where both the \skl stabilization mechanisms and magnetic ground states remain controversial. 
We address these here by investigating both the static and dynamical spin properties of the \centrosymmetric \skl host \gdpdsi using \musrLongSpecFirst. 
We find that spin fluctuations in the \noncoplanar \skl phase are highly anisotropic, implying that spin anisotropy plays a prominent role in stabilizing this phase. 
We also observe strongly-anisotropic spin dynamics in the ground-state (\icOne) \incommensurate magnetic phase of the material, indicating that it hosts a meron-like multi-$q$ structure. 
In contrast, the higher-field, coplanar \icTwo phase is found to be single-$q$ with nearly-isotropic spin dynamics. 
\end{abstract}

\maketitle

\renewcommand{\figurename}{\bf Fig.} 
\renewcommand{\tablename}{\bf Table} 

\renewcommand\thetable{\arabic{table}}


Topological spin textures can support exotic spin dynamics with a range of potential applications~\cite{li2023magnetic,khatua2023experimental}. 
Especially promising are materials hosting skyrmions, which are topologically-protected, \noncoplanar vortices in the magnetization, that behave as extended particles~\cite{lancaster2019skyrmions,li2023magnetic}. 
While in bulk they are usually found in \noncentrosymmetric materials and stabilized by \dmiLongFirst, which select a skyrmion helicity~\cite{morikawa2013crystal}, they were also found to be stabilized by competing magnetic interactions in bulk \centrosymmetric compounds with no preferred helicity and no net \dmi~\cite{li2023magnetic,khatua2023experimental}. 
Examples are \gdpdsi, with a triangular spin lattice~\cite{kurumaji2019skyrmion,hirschberger2020high,hirschberger2020topological,zhang2020anomalous,spachmann2021magnetoelastic,paddison2022magnetic,ju2023polarized,kolincio2023kagome,nakajima2024polarized,dong2024fermi,mayoh2024crystal}, 
\gdrual with a breathing kagome spin lattice~\cite{hirschberger2019skyrmion,hirschberger2021nanometric,kolincio2023kagome,ogunbunmi2023magnetic}, and 
\gdrusi~\cite{khanh2020nanometric,yasui2020imaging,khanh2022zoology,matsuyama2023quantum,wood2023double,eremeev2023insight,spethmann2024sp,dong2023magnetic,paddison2024spin,spethmann2024sp,huddart2024field}, 
\gdruge~\cite{yoshimochi2024multistep}, and 
\eual~\cite{takagi2022square,shang2021anomalous,zhu2022spin} with tetragonal spin lattices. 
Common to these highly-symmetric rare-earth materials is that they host \incommensurate \sklLongAdjFirst phases with very small ($1.9$--$\SI{3.5}{\nano\meter}$) skyrmions that are stable under higher applied fields ($\SI{\sim 1}{\tesla}$) and over a wider range of $T$ than their \dmi-stabilized counterparts~\cite{li2023magnetic,khatua2023experimental}. 

The stabilization mechanism for these \centrosymmetric skyrmions is controversial, with suggestions including: 
(i) short-range geometrical frustration~\cite{okubo2012multiple,leonov2015multiply,lohani2019quantum,hirschberger2021nanometric,pang2023unravelling}, 
(ii) long-range \rkkyLongFirst interactions plus 
dipolar~\cite{bouaziz2022fermi,paddison2022magnetic,eremeev2023insight,matsuyama2023quantum,dong2023magnetic,dong2024fermi} or 
biquadratic exchange~\cite{ozawa2017zero,hayami2017effective,hayami2021square,hayami2022multiple,khanh2022zoology,hayami2023anisotropic,yoshimochi2024multistep,paddison2024spin}, 
and (iii) competition of orbital-dependent exchange~\cite{nomoto2020formation,nomoto2023ab}. 
Most, but not all~\cite{ozawa2017zero,hayami2017effective}, such scenarios require spin anisotropy. 
The \zfLongAdjFirst ground state is also contentious~\cite{ju2023polarized}, with early studies of \gdpdsi suggesting an exotic, triple-$q$ magnetic structure~\cite{kurumaji2019skyrmion,hirschberger2020high} 
[\eg, a lattice of merons and antimerons (\mamlAlt), which are half-skyrmion-like spin textures~\cite{yu2018transformation,khanh2022zoology,yoshimochi2024multistep,spethmann2024sp}], 
while recent calculations under scenario (ii) indicated a simpler, single-$q$ helical ground state~\cite{bouaziz2022fermi,paddison2022magnetic}, as found in \dmi-stabilized \skl hosts~\cite{kanazawa2017noncentrosymmetric}. 
Investigations of \centrosymmetric \skl hosts have focused on their static and topological properties, with less attention~\cite{zhu2022spin} paid to emergent spin dynamics~\cite{leonov2015multiply}, which can elucidate both the stabilization mechanism and single- or multi-$q$ structure. 
Characteristic dynamics of \skl and related spin textures have been observed in a range of \dmi-stabilized \skl hosts~\cite{khatua2023experimental,li2023magnetic,garst2017collective,weber2022topological,soda2023asymmetric}. 
For accessing these, \musrLongSpecFirst, a local-probe technique sensitive to spin fluctuations over a unique frequency range (applicable also at low applied fields and to conductive samples)~\cite{blundell2021muon}, has proven valuable~\cite{franke2018magnetic,stefancic2018origin,hicken2020magnetism,hicken2021megahertz,wilson2021spin,hicken2022magnetism,hicken2022depth}. 
However, the narrow $T$ range of \dmi skyrmions~\cite{li2023magnetic,khatua2023experimental,kanazawa2017noncentrosymmetric} limited the ability to extract the underlying spin-wave dispersions unambiguously. 
Namely, when $T$ is comparable to the transition temperature $T_N$, spin dynamics can become dominated by multi-magnon processes and/or critical fluctuations (usually over $T_N/2 \lesssim T \lesssim 2 T_N$), obscuring the underlying spin-wave dispersion. 

\begin{figure}[!t] 
\centering
\includegraphics[width=1\columnwidth]{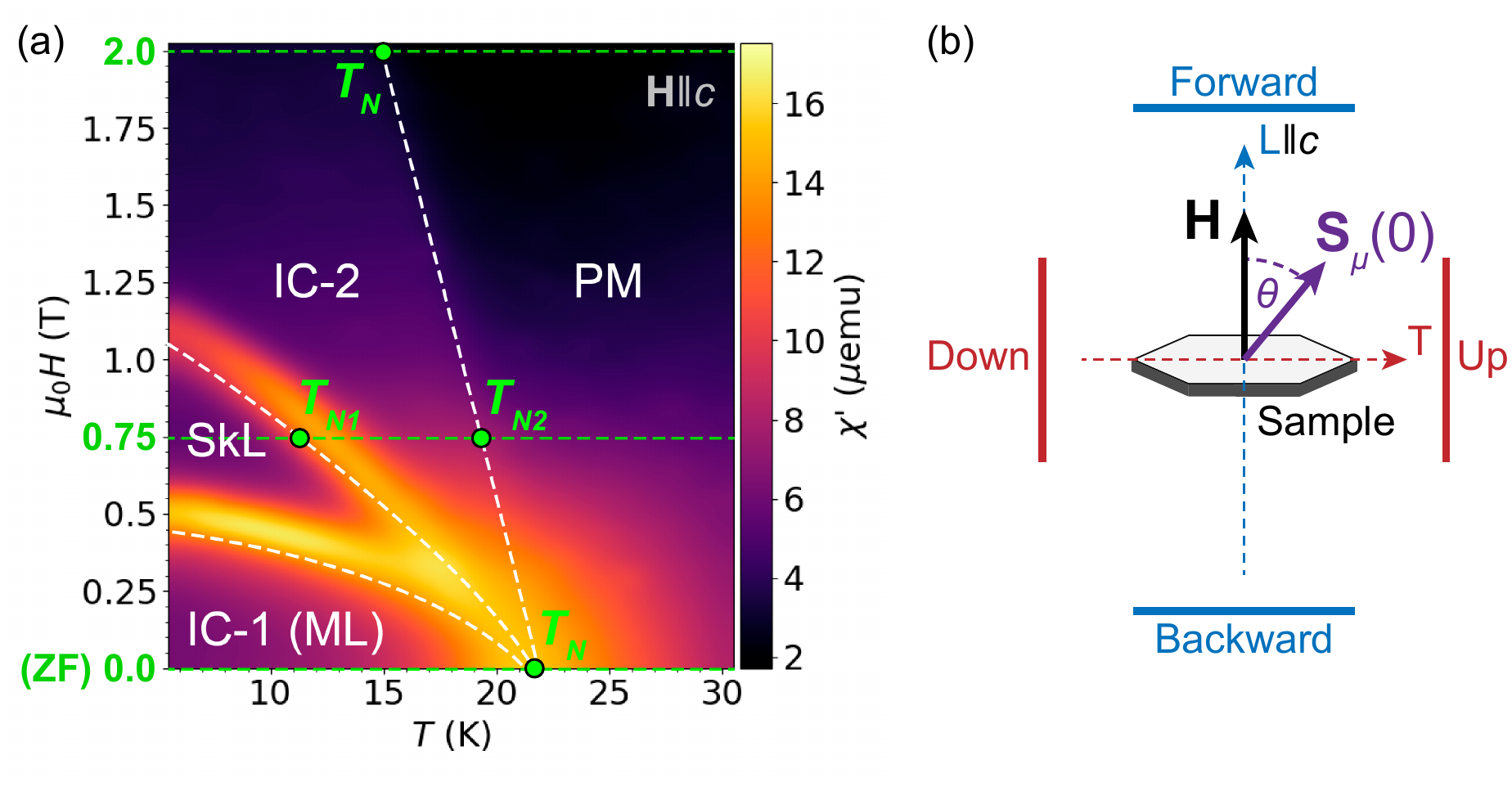}
\caption{%
(a) Real part of AC susceptibility with applied field along the 
$c$ axis. 
Phase boundaries (white) and \musr scans (green) with the corresponding transition temperatures $T_N$ are shown. 
Phase assignments, including our suggested assignment of the \icOne phase as a \mamlAltLongFirst, combine evidence from a range of techniques~\cite{kurumaji2019skyrmion,hirschberger2020high,hirschberger2020topological}. 
(b) Sample and detector arrangement for \musr measurements. 
The initial muon spin $\myvec{S}_\mu(0)$ lies at an angle $\theta$ to the applied field $\myvec{H} \parallel c$, which points along the longitudinal direction \lDir (blue). 
The transverse direction \tDir (red) lies in the hexagonal $ab$ plane.%
}
\label{fig_phasediag_acsusc_musr_setup}
\end{figure}
In this Letter, we investigate the most-studied \centrosymmetric \skl host \gdpdsi~\cite{kurumaji2019skyrmion} using \musr and AC susceptibility, complemented by \dftLongFirst calculations of muon stopping sites~\cite{supp}. 
We find clear signatures of spin reorientation transitions between the \incommensurate magnetic phases and reveal the highly anisotropic character of their spin fluctuations. 
By exploiting the relatively wide region of stability of these phases, we characterize the low-energy dispersion relations of their spin-wave excitations. 
The excitations in the high-field coplanar \incommensurate \icTwo phase are consistent with a single-$q$ fan-like spin texture~\cite{kurumaji2019skyrmion,hirschberger2020high} with nearly-isotropic spin fluctuations. 
However, in the ground-state \zf \incommensurate \icOne phase, we find that low-energy in-plane ($ab$-plane) spin fluctuations dominate, while out-of-plane ($c$-axis) fluctuations are almost completely suppressed. 
This, combined with previous \rxsLongFirst and resistivity data~\cite{kurumaji2019skyrmion,hirschberger2020high}, indicates that the \icOne phase is a complex, triple-$q$ magnetic structure, not the recently-predicted single-$q$ helical structure~\cite{bouaziz2022fermi,paddison2022magnetic}.
This disfavours \skl-stabilization scenario (ii) above. 
Finally, in the \skl phase, out-of-plane fluctuations dominate instead, with in-plane fluctuations suppressed as a power-law in $T$. 
We, therefore, argue that spin anisotropy is a key ingredient in stabilizing the \skl phase in \centrosymmetric rare-earth magnets. 


Single crystals of \gdpdsi were synthesized, with five high-quality $\SI{{\approx}9}{\milli\meter\squared} \times \SI{0.6}{\milli\meter}$ plate-like crystallites with $c$-axis normals extracted~\cite{supp,mayoh2024crystal}. 
AC magnetic susceptibility measurements reproduced the phase diagram from previous studies~\cite{kurumaji2019skyrmion,hirschberger2020high,hirschberger2020topological}, including the \skl phase [\cref{fig_phasediag_acsusc_musr_setup}(a)]. 
For \musr measurements, the crystallites were assembled in a mosaic with coaligned $c$-axes~\cite{supp}. 
The initial muon spin pointed along the out-of-plane ($c$-axis) direction in \lfLongAdjFirst measurements ($\theta = 0$), and at an angle of $\theta \approx \SI{50}{\degree}$ from the $c$ axis in \zf and \tfLongAdjFirst measurements [\cref{fig_phasediag_acsusc_musr_setup}(b)]. 
In a \musr experiment, longitudinal- ($\lDirEq \parallel c$) and transverse- ($\tDirEq \perp c$) muon spin components evolve independently~\cite{blundell2021muon}, producing asymmetries $A_j(t) \propto \braket{S_\mu^j(t)}$, where $S_\mu^j(t)$ is the muon spin component $j$ at time $t$, in detector pairs positioned along directions $j = \lDirEq$ and $\tDirEq$ [\cref{fig_phasediag_acsusc_musr_setup}(b)]. 
An intermediate value of $\theta$ in \zf and \tf experiments thus allowed us to track the different impacts of the material's magnetic state on the time evolution of the \lDir and \tDir muon spin components 
from a single experimental run. 
On the other hand, in \lf experiments (where $\theta = 0$) only the \lDir muon spin component could be measured. 
All measurements were made after first \zfcIngLongFirst the sample to base $T$, while magnetic fields were always applied along the $c$-axis, which is an axis of \threefold crystallographic symmetry and thus a magnetic eigenaxis. 

\begin{figure}[!t] 
\centering
\includegraphics[width=0.85\columnwidth]{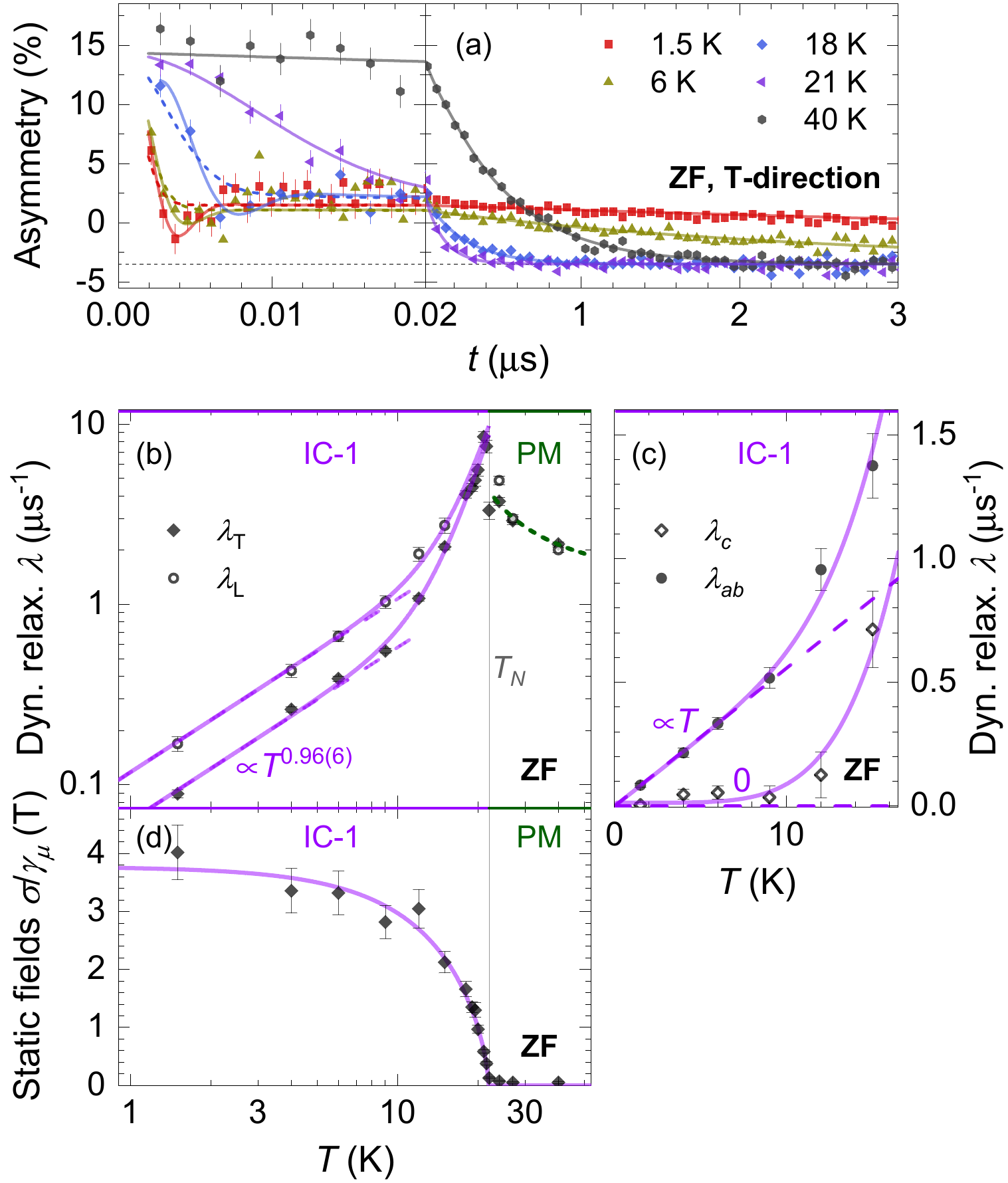}
\caption{%
\icOne phase in \zf. 
(a) $ab$-plane (\tDir-direction) muon asymmetry at early (left) and late times (right). 
Solid lines on the left panel are guides to the eye; dashed lines are fits using the model described in the text. 
Horizontal dashed line shows the background level $a_\mathrm{bgd}$. 
(b) Dynamical relaxation rates and (c) contributions to these rates due to out-of-plane ($c$-axis) and in-plane ($ab$-plane) spin fluctuations. 
Solid lines at $T < T_N$ are guides to the eye, which include a critical divergence near $T_N$ (evident as an increase in $\lambda$ above $\SI{\sim 10}{\kelvin}$); 
dashed lines are the single-power-law low-$T$ limits.
(d) \Quasistatic magnetic fields from \tDir-direction data on panel (a).%
}
\label{fig_icone_tscan}
\end{figure}
We first performed \musr measurements on warming from the \icOne ground-state of \gdpdsi in \zf on the \instrGPS instrument at the Swiss Muon Source (S$\mu$S). 
At low $T$, a highly-damped oscillation in the \tDir-direction ($ab$-plane) muon-spin component was observed at early times $t \ll \SI{0.1}{\micro\second}$ [\cref{fig_icone_tscan}(a)] due to a broad distribution of \quasistatic local fields at the muon site~\cite{supp} that originate from long-range \incommensurate magnetic ordering of \gdIon spins. 
At later times, up to $t \approx \SI{10}{\micro\second}$, a further exponential relaxation was observed due to dynamical fluctuations of \gdIon spins slower than the muon precession frequency~\cite{blundell2021muon}. 
The measured \tDir-direction asymmetry data were fitted using 
$A(t) = \left[ a_s - a_r \right] e^{-\sigma^2 t^2} + a_r e^{-\lambda t} + a_\mathrm{bgd}$, 
where $a_s = \const$ is the total sample asymmetry and $a_\mathrm{bgd} = \const$ is the background due to muons hitting the sample holder. 
The early-time damping rate $\sigma$ in this model is proportional to the average strength of \quasistatic local magnetic fields at the muon site~\cite{supp} 
and is expected to roughly scale with the magnitude of ordered moments in the sample. 
$a_r$ is the late-time relaxing asymmetry due to the fraction of local fields that initially point along the measured muon spin component~\cite{supp}, and
$\lambda$ is the dynamical relaxation rate due to slow fluctuations of fields orthogonal to the measured muon-spin component~\cite{supp,blundell2021muon}. 

Fit results for the \tDir direction are shown in \cref{fig_icone_tscan}(b,d) with the transition temperature $T_N = \SI{22(1)}{\kelvin}$ consistent with AC susceptibility [\cref{fig_phasediag_acsusc_musr_setup}(a)]. 
In the itinerant \paramagLongFirst regime at $T > T_N$ 
we find $\sigma \approx 0$. 
Around $T \approx T_N$, critical spin fluctuations cause $\lambda_{\mathrm{T}}$ to exhibit a divergence characteristic of a continuous phase transition as we enter the \icOne phase. 
In the ordered \icOne phase ($T < T_N$) the average local-field strength increases and saturates as an order parameter, $\sigma \propto [1 - (T/T_N)^{3/2}]^\beta$~\cite{blundell2021muon,zhu2022spin} with $\beta = 0.7(1)$, which is large but close to $\beta = 0.50(5)$ found in the \centrosymmetric \skl host \eual in \zf~\cite{zhu2022spin}. 
At low $T$, slow spin fluctuations cross over into a power-law $\lambdaTDirEq \propto T^p$ with $p = 0.96(6)$. 
A low-$T$ power-law dependence of the dynamical relaxation rate could be understood within spin-wave theory for a two-magnon process, which for a single gapless magnon band predicts~\cite{beeman1968nuclear,jansa2018observation,supp} 
\begin{equation}
p = \frac{2 D}{s} - 1 ,
\label{eq_sw_power}
\end{equation}
where $D \leq 3$ is the (integer) dimensionality of spin-wave excitations and $s$ is the dominant power in their dispersion relation $\omega \propto |\myvec{q} - \myvec{q_0}|^s$ around the ordering \wavevector $\myvec{q_0}$. 
Usually, $s = 2$ for ferromagnetic ($\myvec{q_0} = 0$) and $s = 1$ for antiferromagnetic and \incommensurate states ($\myvec{q_0} \neq 0$)~\cite{jensen1991rare}. 
The measured \tDir-direction $p \approx 1$ in the \icOne phase would thus correspond to a \oneDim (single-$q$) magnetic structure ($D = s = 1$), if the single-band approximation were valid. 

To test this, we also fitted the \lDir-direction ($c$-axis) data using the same model to obtain the relaxation rate \lambdaLDir [\cref{fig_icone_tscan}(b)]. 
Assuming bulk uniaxial symmetry, 
we have~\cite{blundell2021muon,supp} 
$\lambdaLDirEq = 2 \lambda_{ab}$ and $\lambdaTDirEq = \lambda_{ab} + \lambda_{c}$, 
where $\lambda_{ab}$ and $\lambda_{c}$ are the relaxation contributions 
due to dynamical fluctuations of in-plane and out-of-plane magnetic fields at the muon site, respectively. 
At the nearly-symmetric in-plane muon site that we obtain from \dft stopping site calculations (near the centre of a \gdIon triangle)~\cite{supp}, $\lambda_{ab}$ and $\lambda_{c}$ ultimately arise from fluctuations of in-plane and out-of-plane \gdIon spin components, respectively. 
This can be seen from a 
symmetry %
decomposition of long-\wavelength \gdIon spin textures under local $ab$-plane reflections and 
$c$-axis rotations (for details, see the \suppCite). 
\cref{fig_icone_tscan}(c) shows the extracted $\lambda_{ab}$ and $\lambda_{c}$. 
Remarkably, in contrast to the single-magnon-band approximation where we should have $\lambda_{ab} \propto \lambda_c \propto T^p$ (for a proof, applicable also to general single-$q$ states, see the \suppCite), we instead find $\lambda_{ab} \propto T$ but $\lambda_{c} \approx 0$
in the \icOne phase. 
Our first result is, therefore, that the \icOne phase is not a simple single-$q$ magnetic structure, as was predicted~\cite{bouaziz2022fermi,paddison2022magnetic}. 
Instead, there appear to be multiple, highly-anisotropic spin-fluctuation modes in this phase. Such behavior is expected for extended, multi-$q$ spin textures~\cite{leonov2015multiply,garst2017collective}, such as the hypothesized \mamlAlt state~\cite{kurumaji2019skyrmion,hirschberger2020high}. 
Their predominantly in-plane nature appears consistent with easy-plane anisotropy found in this phase~\cite{ju2023polarized}. 

\begin{figure}[!t] 
\centering
\includegraphics[width=0.77\columnwidth]{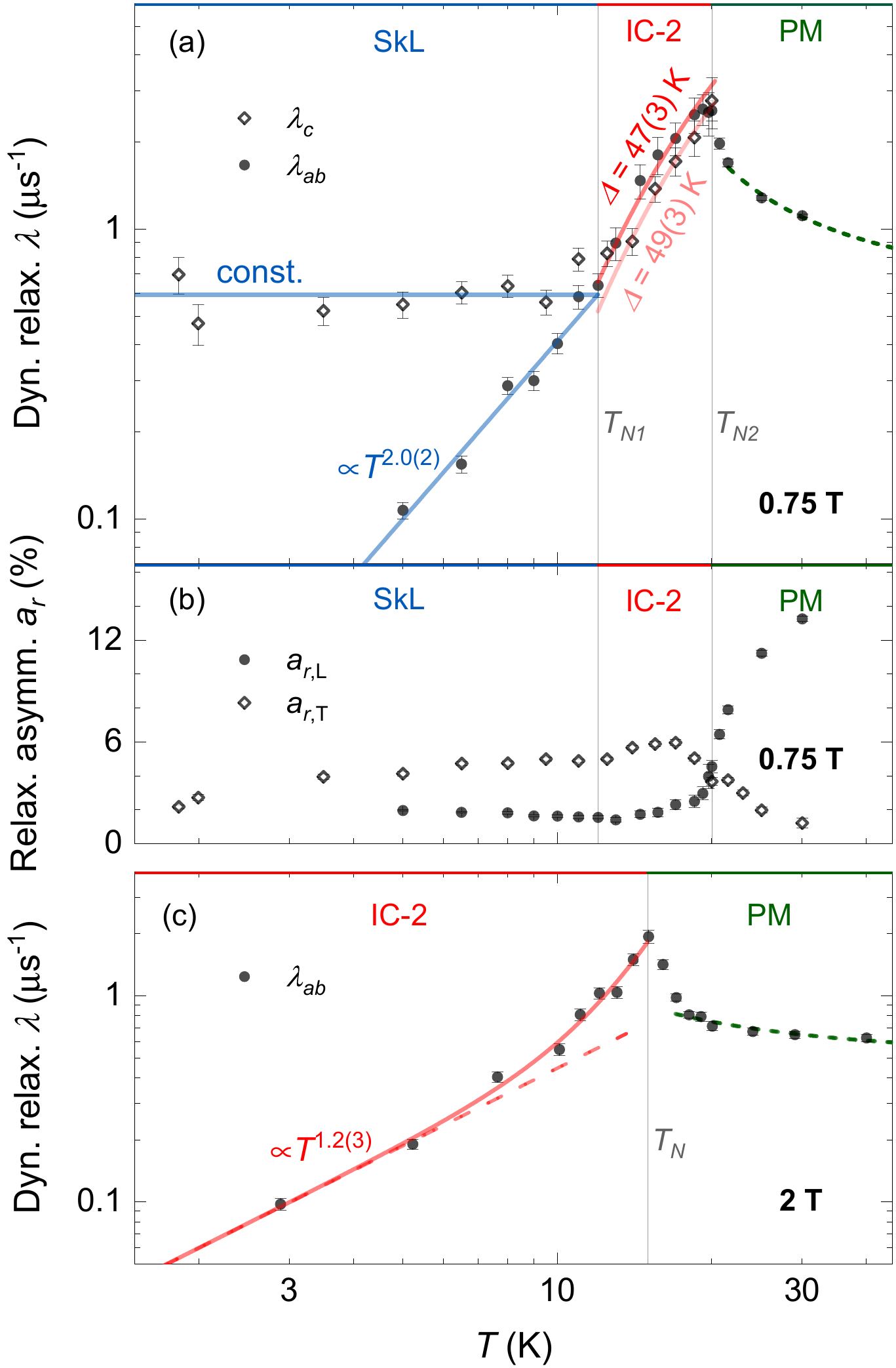}
\caption{%
(a) Contributions to relaxation due to out-of-plane ($c$-axis) and in-plane ($ab$-plane) spin fluctuations in the \skl and \icTwo phases in a $\SI{0.75}{\tesla}$ applied field. 
(b) Relaxing asymmetry due to \quasistatic local fields along the $c$ axis (L) and within the $ab$ plane (T) under these conditions. 
(c) Relaxation due to in-plane fluctuations in the \icTwo phase in a $\SI{2}{\tesla}$ field.%
}
\label{fig_skl_ictwo_tscan}
\end{figure}
Next, we turn to the \icTwo and \skl phases. 
Here, we performed separate \musr measurements in a \tf of $\SI{0.75}{\tesla}$ on the \instrGPS instrument at S$\mu$S and a \lf of $\SI{0.75}{\tesla}$ on the \instrHiFi instrument at the STFC-ISIS Facility. 
We warmed from the low-$T$ \skl to the intermediate \icTwo phase, and finally to the \paramag phase [\cref{fig_phasediag_acsusc_musr_setup}(a)]. 
The late-time data were well described by the same model as for the \zf data, simplified to $A(t) = a_r e^{-\lambda t} + a_\mathrm{bgd}$ at late times. 
From \tDir-direction \tf data and \lDir-direction \lf data we again extracted in-plane $\lambda_{ab}$ and out-of-plane $\lambda_{c}$ spin-fluctuation contributions.

The resulting relaxation rates are shown in \cref{fig_skl_ictwo_tscan}(a) with transition temperatures $T_{N1} = \SI{12(1)}{\kelvin}$ and $T_{N2} = \SI{20(1)}{\kelvin}$ consistent with AC susceptibility [\cref{fig_phasediag_acsusc_musr_setup}(a)]. 
The in-plane relaxation $\lambda_{ab}$ shows a critical divergence at $T \approx T_{N2}$ due to a continuous transition between the \paramag and \icTwo phases. 
In the \icTwo regime ($T_{N1} < T < T_{N2}$) we find nearly-isotropic spin fluctuations with $\lambda_{c} \approx \lambda_{ab}$. 
(Fitting these to a gapped model ${\propto}e^{-\Delta/T}$ also yields the same characteristic energy scales $\Delta = 47(3)$ and $\SI{49(3)}{\kelvin}$ for $\lambda_{c}$ and $\lambda_{ab}$, respectively.) 
Any low-$T$ power-law spin-wave behaviour is masked by near-critical fluctuations and interrupted by a transition to the \skl phase at $T_{N1}$. 

The phase transition at $T_{N1}$ does not show any critical divergence in the muon relaxation rate, only a change of slope, implying that it is first-order, consistent with the topologically-\nontrivial nature of the \skl~\cite{leonov2015multiply,li2023magnetic}. 
While nearly-isotropic at $T_{N1}$, 
we find that highly-anisotropic spin fluctuations 
emerge 
with lowering $T < T_{N1}$, 
similarly to the \icOne phase. 
However, while in the \icOne phase spin fluctuations were predominantly in-plane ($\lambda_c \ll \lambda_{ab}$) the dominant spin fluctuations in the \skl phase are instead out-of-plane, with $\lambda_c \approx \const \gg \lambda_{ab} \propto T^p$ and $p = 2.0(2)$. 
We note that $p \approx 2$ would uniquely correspond to $D = 3$, $s = 2$ (\threeDim ferromagnetic) spin excitations under the single-band spin-wave approximation [\cref{eq_sw_power}], but this is inconsistent with a $T$-independent $\lambda_c$~\cite{supp}. 
Instead, there appears to be a large low-energy spin \dosLong due to multiple spin fluctuation modes, as expected for \skl phases~\cite{mochizuki2012spin,leonov2015multiply,garst2017collective,weber2022topological,soda2023asymmetric}, that are predominantly polarized out-of-plane (\eg, skyrmion breathing modes~\cite{mochizuki2012spin,garst2017collective}). 

We next turn to static properties of the \skl and \icTwo phases in an applied field of $\SI{0.75}{\tesla}$. 
\cref{fig_skl_ictwo_tscan}(b) shows late-time relaxing asymmetries \arTDir and \arLDir in the in-plane (\tDir)  and out-of-plane (\lDir)  directions. 
Both change rather abruptly at $T \approx T_{N2}$ due to the onset of magnetic order. 
Assuming bulk uniaxial symmetry
we expect~\cite{supp,blundell2021muon}
$\arTDirEq \propto \braket{\hat{B}_{a}^2} = \braket{\hat{B}_b^2}$ and $\arLDirEq \propto \braket{\hat{B}_c^2}$, 
where $\myvec{\hat{B}} = \myvec{B}/|\myvec{B}|$ is the initial direction of a \quasistatic local field $\myvec{B}$ at the muon site. 
In \cref{fig_skl_ictwo_tscan}(b) we see that $\arTDirEq$ exhibits a broad peak in the \icTwo phase at $T_{N1} < T < T_{N2}$, while $\arLDirEq$ exhibits a minimum. 
These both indicate approximately 
coplanar \quasistatic magnetism in this phase. 
In the \skl phase, the difference between $\arTDirEq$ and $\arLDirEq$ becomes smaller, implying that local field directions become more isotropic, as expected for a \noncoplanar spin texture~\cite{li2023magnetic,khatua2023experimental,lancaster2019skyrmions}. 
Our observation of a coplanar magnetism in the \icTwo phase and \noncoplanar magnetism in the \skl phase is consistent with \rxs results~\cite{kurumaji2019skyrmion}, where this was argued based on the ellipticity 
of the magnetic moments 
of individual magnetic \bragg peaks. 
%
However, our conclusions are based on a complementary~\cite{ju2023polarized}, real-space determination of local field directions only accessible to local probes like the muon~\cite{blundell2021muon}. 
Intriguingly, \via field-dependent \quasistatic~\musr measurements, we find that the width of the local field distribution does not scale with the average local field strength in the \skl phase, but does do so in the \icOne and \icTwo phases (see the \suppCite). %

Finally, to assess the low-$T$ dynamics of the \icTwo phase, we performed \musr measurements in a \lf of $\SI{2}{\tesla}$ on the \instrHiFi instrument. 
The late-time \lDir-direction muon data were well described by the same model as for the $\SI{0.75}{\tesla}$ data, where the fitted relaxation rate $\lambdaLDirEq = 2 \lambda_{ab}$ [\cref{fig_skl_ictwo_tscan}(c)] shows a transition temperature $T_N = \SI{15(1)}{\kelvin}$ consistent with AC susceptibility [\cref{fig_phasediag_acsusc_musr_setup}(a)]. 
Near $T \approx T_N$ the muon relaxation rate exhibits a critical divergence characteristic of a continuous phase transition as we enter the \icTwo phase, and at $T < T_N$ it crosses over into a low-$T$ power-law $\lambdaLDirEq \propto T^p$ with $p = 1.2(3)$. 
The observed isotropy of spin fluctuations in this phase [\cref{fig_skl_ictwo_tscan}(a)] makes single-band spin-wave theory [\cref{eq_sw_power}] applicable, with the measured $p \approx 1$ indicating that the \icTwo phase is a \oneDim (single-$q$) magnetic structure ($D = s = 1$), as previously suggested~\cite{kurumaji2019skyrmion,hirschberger2020high}. 
This stands in contrast to multi-$q$ \icOne and \skl phases found at lower applied fields. 


To summarize, our finding of different low-$T$ in-plane and out-of-plane spin fluctuations in both 
\noncoplanar %
\skl and ground-state \icOne phases of \centrosymmetric \gdpdsi contrasts with isotropic fluctuations found in the 
coplanar %
single-$q$ \icTwo phase. 
This should supply a clue to the stabilization mechanism for \centrosymmetric skyrmions. 
In contrast to \dmi-stabilized skyrmions~\cite{mochizuki2012spin,garst2017collective,xing2020magnetic,weber2022topological,soda2023asymmetric}, systematic predictions for \skl spin dynamics for different stabilization-mechanisms are lacking.
Nevertheless, it seems unlikely that a spin model without strong intrinsic anisotropy could explain the observed highly-anisotropic \skl and \icOne spin dynamics. 
This agrees with suggestions that spin anisotropy~\cite{leonov2015multiply,ju2023polarized,hayami2023anisotropic}, combined with long-range interactions, is important for stabilizing the \skl phase~\cite{paddison2022magnetic,hayami2021square,paddison2024spin,hirschberger2021nanometric}. 
A quantitative determination of the anisotropy of intrinsic magnetic interactions in \gdpdsi would thus be crucial~\cite{ju2023polarized}. 
We note that anisotropic magnetic interactions are also found in other representatives of the hexagonal \rpdsi family ($R$ = rare earth)~\cite{frontzek2009magnetic,frontzek2006magneto}, in the tetragonal \centrosymmetric \skl hosts \gdrusi and \gdruge~\cite{sarkar2024unveiling,nomoto2023ab,khanh2020nanometric,garnier1996giant,garnier1995anisotropic}, and even (weakly) in cubic \gdAtom-based magnets~\cite{jia2007nearly}. 
Furthermore, our observation that the ground-state \icOne phase is a multi-$q$ magnetic structure (triple-$q$, based on previous \rxs~\cite{kurumaji2019skyrmion,hirschberger2020high} and neutron-scattering~\cite{nakajima2024polarized} \bragg-peak studies) suggests it is the exotic \mamlAlt state, as hypothesized in Ref.~\cite{kurumaji2019skyrmion} from \rxs and 
resistivity measurements. 
While double-$q$ (square-lattice) \mamlAlt-like states have recently been reported in thin-plate \dmi-based \coznmn~\cite{yu2018transformation}, and bulk tetragonal \centrosymmetric \gdruge~\cite{yoshimochi2024multistep} and \gdrusi~\cite{khanh2022zoology} (in two phases), the \icOne phase would be a unique example of a triple-$q$ magnetic \mamlAlt. 
Furthermore, the \icOne phase is the \zf ground state in \gdpdsi, while the \mamlAlt states in \coznmn and \gdruge are only stabilized under an applied field, 
and the claimed \zf \mamlAlt phase in \gdrusi~ has recently been reinterpreted as a topological-charge-stripe state instead~\cite{wood2023double,spethmann2024sp,huddart2024field}. 
The multi-$q$ nature of the \zf \icOne phase 
represents another challenge to theory, as calculations suggested that the \zf state should instead be single-$q$ under the {\rkky}+dipolar skyrmion stabilization scenario (ii)~\cite{bouaziz2022fermi,paddison2022magnetic}. 
Our findings thus disfavour this as the skyrmion stabilization mechanism. 
Nevertheless, \rkky interactions were found to be strong in the related \centrosymmetric \skl host \gdrusi via \arpesLongFirst~\cite{eremeev2023insight,dong2023magnetic} and quantum-oscillation measurements~\cite{matsuyama2023quantum}, complemented by \abinitio calculations~\cite{khanh2022zoology,matsuyama2023quantum,eremeev2023insight,dong2023magnetic}, so they could still play a role. 
A very recent \arpes and \abinitio study of \gdpdsi seems to support this~\cite{dong2024fermi}. 


In conclusion, in our muon-spectroscopy study of \centrosymmetric \gdpdsi we have found large anisotropy in spin dynamics, with qualitatively different behaviour of dominant out-of-plane and subdominant in-plane spin fluctuations in the \skl phase. 
We have also established the meron-like triple-$q$ nature of its \incommensurate \icOne ground state with dominant in-plane, and nearly-absent out-of-plane, fluctuations. 
The higher-field \icTwo phase was found to be coplanar and single-$q$ with isotropic spin fluctuations. 
Our results suggest that the enigmatic stabilization mechanism behind \skl phases in \centrosymmetric \gdAtom- and \euAtom-based materials is likely to be intimately related to  spin anisotropy, 
and not solely {\rkky}+dipolar. 
Further local-probe studies of these and related \centrosymmetric compounds~\cite{li2023magnetic,khatua2023experimental,kotsanidis1990magnetic,tang2011crystallographic} should be informative in exploring this. 
Studies of anisotropic spin excitations and their dispersions \via inelastic neutron scattering~\cite{paddison2022magnetic}, and the anisotropy of static spin correlations \via \rxs, especially on single crystals, would also be valuable. 
Finally, our \musr methods could also be extended to study low-$T$ spin anisotropy and dynamics of metastable skyrmions in \dmi-based \skl hosts~\cite{li2023magnetic}. 


Research data and code presented in this paper are available at \url{https://dx.doi.org/10.6084/m9.figshare.28024910}.


We thank Martin Klanjšek and Matej Pregelj at the Jo\v{z}ef Stefan Institute, Slovenia for helpful discussions. 
Parts of the work were carried out at the STFC ISIS Muon Source, United Kingdom and at the Swiss Muon Source, Paul Scherrer Institute, Switzerland and we are grateful for the provision of beamtime and experimental support. 
We thank Raymond Fan and Paul Steadman for enabling us to use the Quantum Design MPMS3 at the I10 support laboratory, Diamond Light Source for AC susceptibility measurements. 
Computing resources were provided by the Durham HPC Hamilton cluster. 
M.G. acknowledges the financial support of the Slovenian Research and Innovation Agency through Program No. P1-0125 and Projects No. Z1-1852, N1-0148, J1-2461, J1-50008, J1-50012, N1-0345, and N1-0356. 
We acknowledge the financial support of the Engineering and Physical Sciences Research Council (EPSRC, UK) through Grants No. EP/N032128/1 and EP/N024028/1. 
The work at the University of Warwick was also funded by EPSRC, UK through Grant No. EP/T005963/1. 

%

\end{document}


\selectlanguage{english}

\preprint{APS/123-QED}

\title{Supplemental Material:\\Anisotropic Skyrmion and Multi-$q$ Spin Dynamics in \Centrosymmetric \gdpdsi}



\author{M. Gomil\v{s}ek} 
\email{matjaz.gomilsek@ijs.si} 
\affiliation{Jo\v{z}ef Stefan Institute, Jamova c.~39, SI-1000 Ljubljana, Slovenia}
\affiliation{Faculty of Mathematics and Physics, University of Ljubljana, Jadranska u.~19, SI-1000 Ljubljana, Slovenia}
\affiliation{Department of Physics, Durham University, South Road, Durham DH1 3LE, United Kingdom}


\author{T. J. Hicken} 
\affiliation{Laboratory for Muon Spin Spectroscopy, Paul Scherrer Institut, 5232 Villigen PSI, Switzerland}
\affiliation{Department of Physics, Durham University, South Road, Durham DH1 3LE, United Kingdom}

\author{M. N. Wilson} 
\affiliation{Department of Physics and Physical Oceanography, Memorial University, A1B 3X7, Canada}
\affiliation{Department of Physics, Durham University, South Road, Durham DH1 3LE, United Kingdom}


\author{K. J. A. Franke}
\affiliation{School of Physics and Astronomy, University of Leeds, LS2 9JT, United Kingdom}
\affiliation{Department of Physics, Durham University, South Road, Durham DH1 3LE, United Kingdom}


\author{B. M. Huddart} 
\affiliation{Clarendon Laboratory, University of Oxford, Department of Physics, Oxford OX1 3PU, United Kingdom}
\affiliation{Department of Physics, Durham University, South Road, Durham DH1 3LE, United Kingdom}


\author{A. \v{S}tefan\v{c}i\v{c}} 
\affiliation{University of Warwick, Department of Physics, Coventry CV4 7AL, United Kingdom}

\author{S. J. R. Holt} 
\altaffiliation[Current addresses: ]{Faculty of Engineering and Physical Sciences, University of Southampton, Southampton SO17 1BJ, United Kingdom}
\altaffiliation{Max Planck Institute for the Structure and Dynamics of Matter, Luruper Chaussee~149, 22761 Hamburg, Germany}
\affiliation{University of Warwick, Department of Physics, Coventry CV4 7AL, United Kingdom}

\author{G. Balakrishnan} 
\affiliation{University of Warwick, Department of Physics, Coventry CV4 7AL, United Kingdom}


\author{D. A. Mayoh} 
\affiliation{University of Warwick, Department of Physics, Coventry CV4 7AL, United Kingdom}


\author{M. T. Birch} 
\affiliation{Max Planck Institute for Intelligent Systems, Heisenbergstrasse~3, D-70569 Stuttgart, Germany}
\affiliation{RIKEN Center for Emergent Matter Science, JP-351-0198 Wako, Japan}
\affiliation{Department of Physics, Durham University, South Road, Durham DH1 3LE, United Kingdom}

\author{S. H. Moody} 
\affiliation{Laboratory for Neutron Scattering and Imaging, Paul Scherrer Institut, 5232 Villigen PSI, Switzerland}
\affiliation{Department of Physics, Durham University, South Road, Durham DH1 3LE, United Kingdom}


\author{H. Luetkens} 
\affiliation{Laboratory for Muon Spin Spectroscopy, Paul Scherrer Institut, 5232 Villigen PSI, Switzerland}

\author{Z. Guguchia} 
\affiliation{Laboratory for Muon Spin Spectroscopy, Paul Scherrer Institut, 5232 Villigen PSI, Switzerland}

\author{M. T. F. Telling} 
\affiliation{ISIS Facility, STFC Rutherford Appleton Laboratory, Didcot, Oxfordshire OX11 0QX, United Kingdom}

\author{P. J. Baker} 
\affiliation{ISIS Facility, STFC Rutherford Appleton Laboratory, Didcot, Oxfordshire OX11 0QX, United Kingdom}


\author{S. J. Clark} 
\affiliation{Department of Physics, Durham University, South Road, Durham DH1 3LE, United Kingdom}

\author{T. Lancaster} 
\affiliation{Department of Physics, Durham University, South Road, Durham DH1 3LE, United Kingdom}


\date{\today}

\maketitle

\renewcommand{\figurename}{\bf Supplementary Fig.}
\crefformat{figure}{#2Supplementary Fig.~#1#3}

\section{Crystal synthesis, \semedx, and \laue \xray measurements}

\begin{figure}[!tb] 
\centering
\includegraphics[width=0.80\columnwidth]{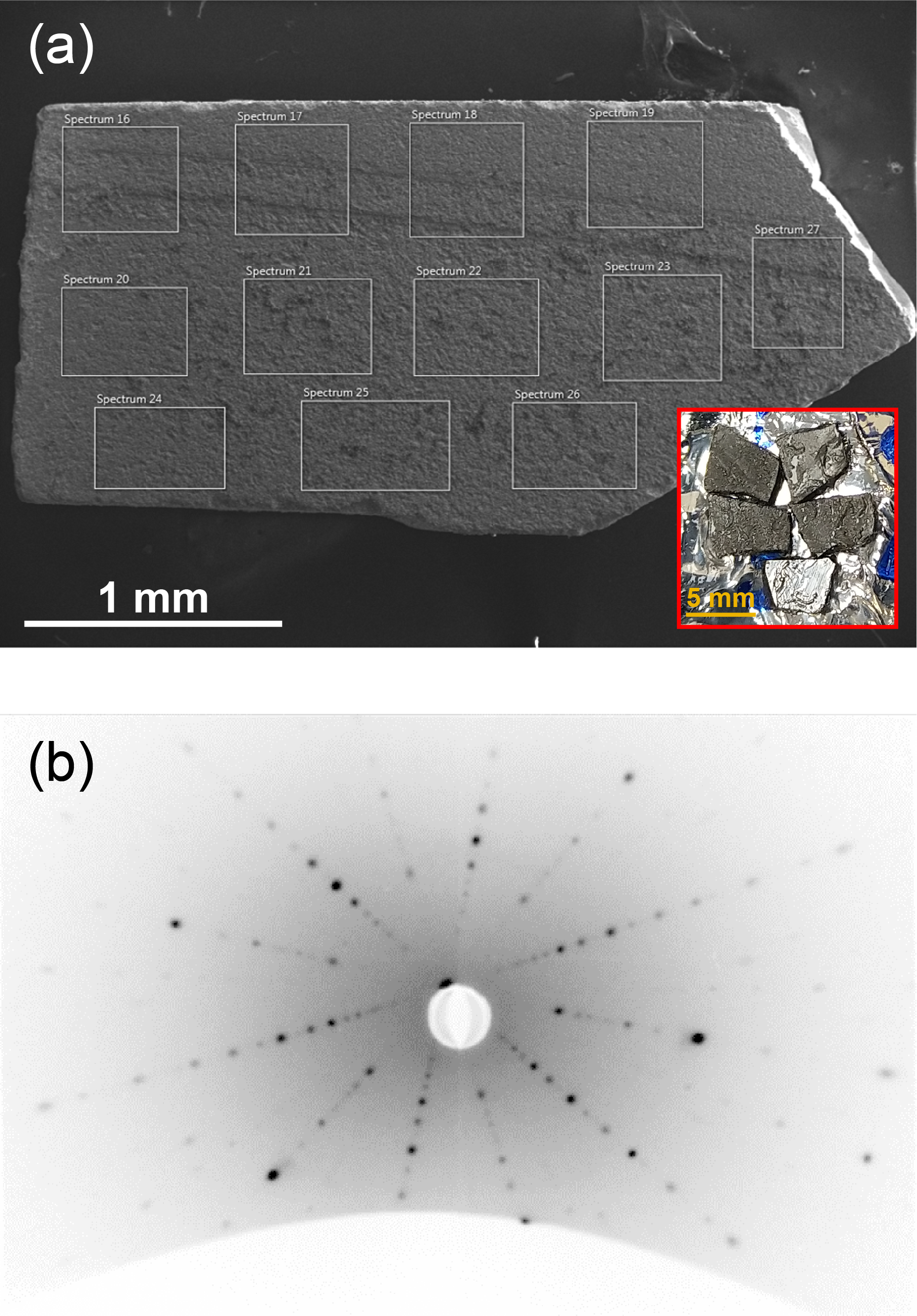}
\caption{%
(a) \sem image of an extracted \gdpdsi single-crystal sample with marked regions from which \edx spectral were analysed. 
Inset: mosaic of five \gdpdsi samples in a silver foil packet for \musr measurements. 
(b) \laue pattern of an isolated single-crystal sample of \gdpdsi.%
}
\label{sfig_crystal_images}
\end{figure}
Single-crystal sample synthesis was performed at the University of Warwick \via the optical floating zone technique~\cite{mayoh2024crystal}. 
Firstly, polycrystalline buttons of \gdpdsi were synthesised by arc melting stoichiometric quantities of \gdAtom ($\SI{99.9}{\percent}$, STREM), \pdAtom ($\SI{99.9}{\percent}$, STREM), and \siAtom ($\SI{99.999}{\percent}$, Sigma-Aldrich). 
The buttons produced were flipped and remelted several times to ensure homogeneity. 
After the polycrystalline buttons were formed, they were recast into a rod for single-crystal growth. 
Single crystals of \gdpdsi were then grown using a four-mirror optical image furnace (equipped with four xenon arc lamps). 
The crystals were grown in a high purity (6N) argon atmosphere. 
Crystal growth was carried out using a range of growth rates, from $8$ to $\SI{12}{\mm/\hour}$, while the feed and seed rods were counter-rotated at speeds ranging from $10$ to $\SI{20}{rpm}$. 
This process yielded a single-crystal boule from which plate-like samples were extracted. 

Compositional analysis was performed \via \semedxLongAnaFirst on $\mathrm{K}_\alpha$ edges of \gdAtom, \pdAtom, and \siAtom on a Zeiss Supra 55-VP FEGSEM [\cref{sfig_crystal_images}(a)]. 
The sample stoichiometry was measured to be \gdpdsiOurSampleWithErr, close to the ideal \gdpdsi. 
Five plate-like, $\SI{{\approx}9}{\milli\meter\squared} \times \SI{0.6}{\milli\meter}$ samples were extracted from the single-crystal boule and polished for \musrLongSpecFirst and AC susceptibility measurements [inset in \cref{sfig_crystal_images}(a)]. 
Sample orientation, with the crystallographic $c$ axis perpendicular to the plates, was checked \via \laue \xray measurements, which found a maximal deviation from perfect alignment below $\SI{4}{\degree}$ and good sample crystallinity [\cref{sfig_crystal_images}(b)]. 

\section{AC susceptibility measurements}

Magnetic susceptibility with magnetic field $\myvec{H} \parallel c$ and a drive field of $\SI{1}{\milli\tesla}$ at $\SI{10}{\hertz}$ was measured using a Quantum Design MPMS3 at the I10 support laboratory, Diamond Light Source. 
The single crystal sample was fixed to a quartz rod with GE varnish. 
The phase diagram [Fig. 1(a) in the main text] was determined using scans of increasing field from $0$ to $\SI{2}{\tesla}$ after \zfcIngLongFirst the sample down from $\SI{30}{\kelvin}$ to each measured temperature. 

\section{\musrLongSpec measurements}

Plate-like samples were assembled in a mosaic before \musr measurements [inset in \cref{sfig_crystal_images}(a)]. 
Muon data from the \instrGPS instrument at the Swiss Muon Source (S$\mu$S) were fit using the \gaussian and exponential models described in the main text, since a damped-cosine-oscillation model [shown as a guide to the eye in Fig. 2(a) in the main text] proved unreliable since only the first minimum of the muon polarization could be clearly resolved at early times due to strong damping of the oscillatory contribution [Fig. 2(a) in the main text], due to the broadness of the \quasistatic local-field distribution at the muon site, which made separating the oscillation frequency from the \gaussian damping rate hard. 
An alternative \kubotoyabe model~\cite{blundell2021muon}, which describes muon polarization under a broad \quasistatic local-field distribution in a powder or an isotropic magnetic system, was not appropriate as our samples were anisotropic single crystals. 
A constant background contribution $a_\mathrm{bgd} = \const$ absorbed the lack of alpha correction of detector efficiencies in some of the datasets [\eg, Fig. 2(a) in the main text], as the two are almost indistinguishable~\cite{blundell2021muon}.  

\subsection{\Quasistatic polarization} 

In ordered magnetic systems, the muon polarization at early times is dominated by \quasistatic local magnetic fields at the muon site. 
Namely, the \quasistatic muon polarization along the detector direction, which we denote $z$, at time $t$ is given by~\cite{blundell2021muon}
\begin{equation}
P(t) = \int \DD{3} \myvec{B} \; p(\myvec{B}) \left[ \cos^2 \theta + \sin^2 \theta \cos(\gamma_\mu B t) \right] ,
\label{eq_quasistatic_polarization}
\end{equation}
where $p(\myvec{B})$ is the probability density function that the muon at its stopping site encounters an initial \quasistatic local magnetic field $\myvec{B}$ (\ie, the field distribution at the muon stopping site), 
$B = |\myvec{B}|$, $\theta$ is the angle between the field $\myvec{B}$ and the detector direction $z$, and $\gamma_\mu = 2 \pi \times \SI{135.53}{\mega\hertz/\tesla}$ is the muon gyromagnetic ratio. 
The muon asymmetry is related to the muon polarization by $A(t) = a_s P(t) + A_\mathrm{bgd}(t)$, where $a_s$ is the total sample asymmetry, and $A_\mathrm{bgd}(t)$ is the background asymmetry due to muons missing the sample and hitting the sample holder [\eg, $A_\mathrm{bgd}(t) = a_\mathrm{bgd} = \const$ in \zfLongFirst]. 

When the field distribution $p(\myvec{B})$ allows for different field magnitudes $B$, the oscillating term in \cref{eq_quasistatic_polarization} averages out to zero at late times, leaving the muons with a finite late-time asymmetry $a_r$ (called the relaxing asymmetry) that can only relax dynamically. 
Its value can be calculated to be 
\begin{equation}
a_r = a_s \braket{\hat{B}_z^2} ,
\label{eq_quasistatic_relaxing_asymmetry}
\end{equation}
where $\myvec{\hat{B}} = \myvec{B}/B$ is the normalized initial \quasistatic local-field direction at the muon site, and $\braket{\cdots}$ is the expectation value under the field distribution $p(\myvec{B})$. 

Often one models the true \quasistatic muon polarization in \cref{eq_quasistatic_polarization} with a damped-cosine-oscillation model
\begin{equation}
P(t) = \left(1 - p_r \right) \cos(\gamma_\mu B^* t) e^{-\sigma^{*2} t^2} + p_r ,
\label{eq_quasistatic_model_polarization}
\end{equation}
where $p_r = a_r / a_s$ is the relaxing polarization [\cref{eq_quasistatic_relaxing_asymmetry}], $B^*$ is the oscillation field (related to the oscillation angular frequency $\omega = \gamma_\mu B^*$), and $\sigma^*$ is the \gaussian damping rate. 
Both this model and the true polarization [\cref{eq_quasistatic_polarization}] have the same initial value $P(0) = 1$ and slope $P'(0) = 0$, while matching their initial second derivatives $P''(0)$ gives 
\begin{equation}
\frac{2 \sigma^{*2}}{\gamma_\mu^2} = \frac{\braket{B_x^2 + B_y^2}}{\braket{\hat{B}_x^2 + \hat{B}_y^2}} - B^{*2} = \frac{\braket{B^2} - \braket{B_z^2}}{1 - \braket{\hat{B}_z^2}} - B^{*2} , 
\label{eq_quasistatic_sigma}
\end{equation}
where $x$ and $y$ are axes perpendicular to the detector direction $z$. 

Fitting the oscillation field $B^*$ in the damped-cosine-oscillation model \cref{eq_quasistatic_model_polarization} gives $B^* \approx \braket{B}$, while the \gaussian model from the main text corresponds to fixing $B^* = 0$. 
In the fitted-damped-cosine-oscillation-model case (with $B^* \approx \braket{B}$), the \gaussian damping rate in \cref{eq_quasistatic_sigma}, which we denote $\sigma' = \sigma^*$, corresponds to a weighted width of the \quasistatic local-field distribution $p(\myvec{B})$ at the muon site. 
For example, for an isotropic local-field distribution (\eg, in a powder), we would obtain $\sigma' \approx (\gamma_\mu/\sqrt{2}) \sqrt{\braket{B^2} - \braket{B}^2}$, \ie, a constant times the \stdevLongAdjFirst width of the distribution of local-field magnitudes~\cite{blundell2021muon}. 
On the other hand, in the \gaussian model from the main text (with $B^* = 0$), the damping rate in \cref{eq_quasistatic_sigma}, which we now denote $\sigma = \sigma^*$, corresponds to a weighted average strength of \quasistatic local fields. 
For example, for an isotropic local-field distribution, we would obtain $\sigma = (\gamma_\mu/\sqrt{2}) \sqrt{\braket{B^2}}$, \ie, the \rmsLongFirst of local-field magnitudes~\cite{blundell2021muon}. 

\subsection{Dynamical relaxation} 

\begin{figure}[!t] 
\centering
\includegraphics[width=0.9\columnwidth]{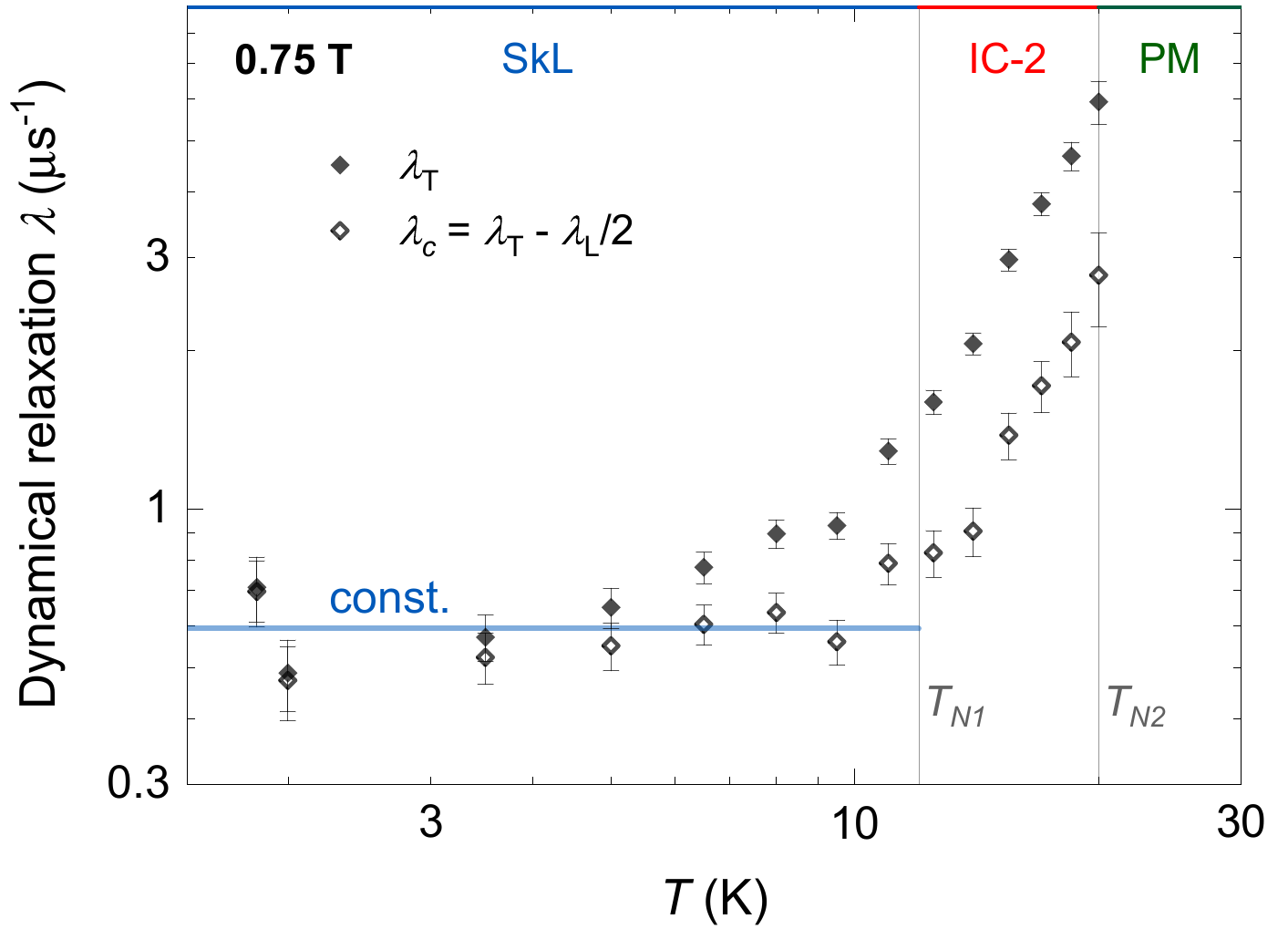}
\caption{%
Comparison of the raw \tDir-direction ($ab$-plane) dynamical muon relaxation rate from \tf measurements (solid symbols) and the extracted relaxation rate due to out-of-plane ($c$-axis) spin fluctuations (unfilled symbols) in the \skl and \icTwo phases in a $\SI{0.75}{\tesla}$ applied field.%
}
\label{fig_skl_raw_tf}
\end{figure}
At late times, exponential decay of muon polarization was observed [Fig. 2(a) in the main text]. 
In general, at late times, dynamical fluctuations of the local magnetic field $\myvec{B}(t)$ at the muon site, which are described by the field--field autocorrelation function $\Phi_{\alpha \beta}(t) = \gamma_\mu^2 \braket{{\delta B}_\alpha(t) {\delta B}_\beta(0)}$ where $\myvec{\delta B}(t) = \myvec{B}(t) - \braket{\myvec{B}}$, induce an exponential decay of the muon polarization function along the detector direction, which we again denote $z$ with perpendicular directions $x$ and $y$, with a relaxation rate given by~\cite{blundell2021muon} 
\begin{equation}
\lambda = \int_0^\infty \cos(\omega_0 t) [ \Phi_{xx}(t) + \Phi_{yy}(t) ] \dd t ,
\label{eq_lambda_field_field}
\end{equation}
where $\omega_0 = \gamma_\mu \braket{B}$ 
is the average \larmor angular frequency of the total magnetic field (applied plus internal). 
In other words, a component ($z$) of the muon spin relaxes at late times due to dynamical fluctuations of the local magnetic field in the plane perpendicular to that component ($xy$-plane). 
If we have a crystal with uniaxial symmetry along the $c$ axis, we thus have relaxation rates $\lambdaLDirEq = 2 \lambda_{ab}$ and $\lambdaTDirEq = \lambda_{ab} + \lambda_{c}$, for muon spin components along $z = \lDirEq \parallel c$ 
and $z = \tDirEq \perp c$ 
directions, respectively, where the individual relaxation-rate contributions are 
\begin{equation}
\begin{aligned}
\lambda_{ab} &= \int_0^\infty \cos(\omega_0 t) \Phi_{aa}(t) \dd t = \lambdaLDirEq/2 ,\\
\lambda_c    &= \int_0^\infty \cos(\omega_0 t) \Phi_{cc}(t) \dd t = \lambdaTDirEq - \lambdaLDirEq/2 ,
\end{aligned}
\label{eq_lambda_dir_contrib}
\end{equation}
and uniaxial symmetry ensures that $\Phi_{aa}(t) = \Phi_{bb}(t)$. 
\cref{fig_skl_raw_tf} shows the fitted \tDir-direction relaxation rate $\lambdaTDirEq$ from \tfLongAdjFirst measurements and the out-of-plane relaxation-rate contribution $\lambda_{c}$ extracted from it \via \cref{eq_lambda_dir_contrib}. 
The latter is also shown on Fig.~3(a) in the main text. 
Where the temperatures at which $\lambdaTDirEq$ and $\lambdaLDirEq$ were measured did not match, $\lambdaLDirEq$ was interpolated to the temperature of a given $\lambdaTDirEq$ measurement before \cref{eq_lambda_dir_contrib} was applied. 

\subsection{Linear spin-wave theory} 

At low $T$, the dynamical relaxation rate under the spin-wave approximation for a single magnon band due to the two-magnon (\raman) process, which is expected the dominant one in relaxing the muon spin~\cite{beeman1968nuclear}, is given by~\cite{beeman1968nuclear,jansa2018observation} 
\begin{equation}
\lambda \propto \int g^2(E) f(E) [1 + f(E)] \dd E \propto T^p e^{-\Delta/T} ,
\label{eq_sw_lambda}
\end{equation}
with $p = 2D/s - 1$ [Eq.~(1) in the main text], where $D \leq 3$ is the (integer) dimensionality of spin-wave excitations with a spin \dosLong $g(E)$ at energy $E \propto \omega$, $s$ is the dominant power in their dispersion relation $E - \kB \Delta \propto |\myvec{q} - \myvec{q_0}|^s$ around the ordering \wavevector $\myvec{q_0}$, $\Delta$ is the magnon gap, $f(E) = 1/[\exp(\beta E) - 1]$ is the \boseeinstein distribution function, $\beta = 1/(\kB T)$, and $\kB$ is the \boltzmann constant. 
Note that \cref{eq_sw_lambda} is formally only valid for $p > 0$~\cite{jansa2018observation}. 
When the magnons are gapless ($\Delta = 0$) we recover power-law behaviour, $\lambda \propto T^p$, from the main text. 

Note that \cref{eq_sw_lambda} is valid in all detector directions, up to a $T$-independent prefactor, \ie, we expect $\lambda_{ab} \propto \lambda_c \propto \lambda$ at low $T$ in the single-magnon-band approximation. 
One can see this by observing that in the low-$T$ linear spin wave theory approximation (where we expand all operators up to second order in magnon creation and annihilation operators), thermal expectation values of all spin observables, including directional relaxation-rate contributions, can only be proportional to a weighted sum of $T$-dependent expectation values of magnon occupation numbers $\braket{n_i(\myvec{q})}(T)$~\cite{toth2015linear}, where $\myvec{q}$ is the \wavevector and $i = 1, \ldots, N$ is the magnon band index. 
Furthermore, only $\myvec{q}$ near the minima of the corresponding lowest magnon bands contribute, as muons are only sensitive to low-frequency spin fluctuations~\cite{blundell2021muon}. 
If there is only a single magnon band, $N = 1$, all spin observables thus become proportional to just one $\braket{n(\myvec{q})}(T)$ over a narrow range of $\myvec{q}$, which implies that, to a good approximation, we must have $\lambda_{ab} \propto \lambda_c$ as a function of $T$. 
Note that the same holds true for general single-$q$ spin structures, despite them appearing to exhibit up to three bands in the laboratory reference frame, as they are just single-magnon-band states from the point of view of a co-rotating reference frame~\cite{toth2015linear}. 
On the other hand, if we have multiple truly distinct magnon bands, as we do in multi-$q$ spin structures, multiple distinct $\braket{n_i(\myvec{q})}(T)$ can contribute with different weights to $\lambda_{ab}$ and to $\lambda_c$. 
It thus becomes possible that $\lambda_{ab} \not\propto \lambda_c$, provided that the dominant $\braket{n_i(\myvec{q})}(T)$ contributions are sufficiently different, \ie, that the spin state is sufficiently anisotropic. 
An example of this is seen in the main text in the anisotropic multi-$q$ \skl and \icOne phases [Figs. 3(a) and 2(c) in the main text, respectively]. 

\subsection{Field dependence}

\begin{figure}[!t] 
\centering
\includegraphics[width=0.9\columnwidth]{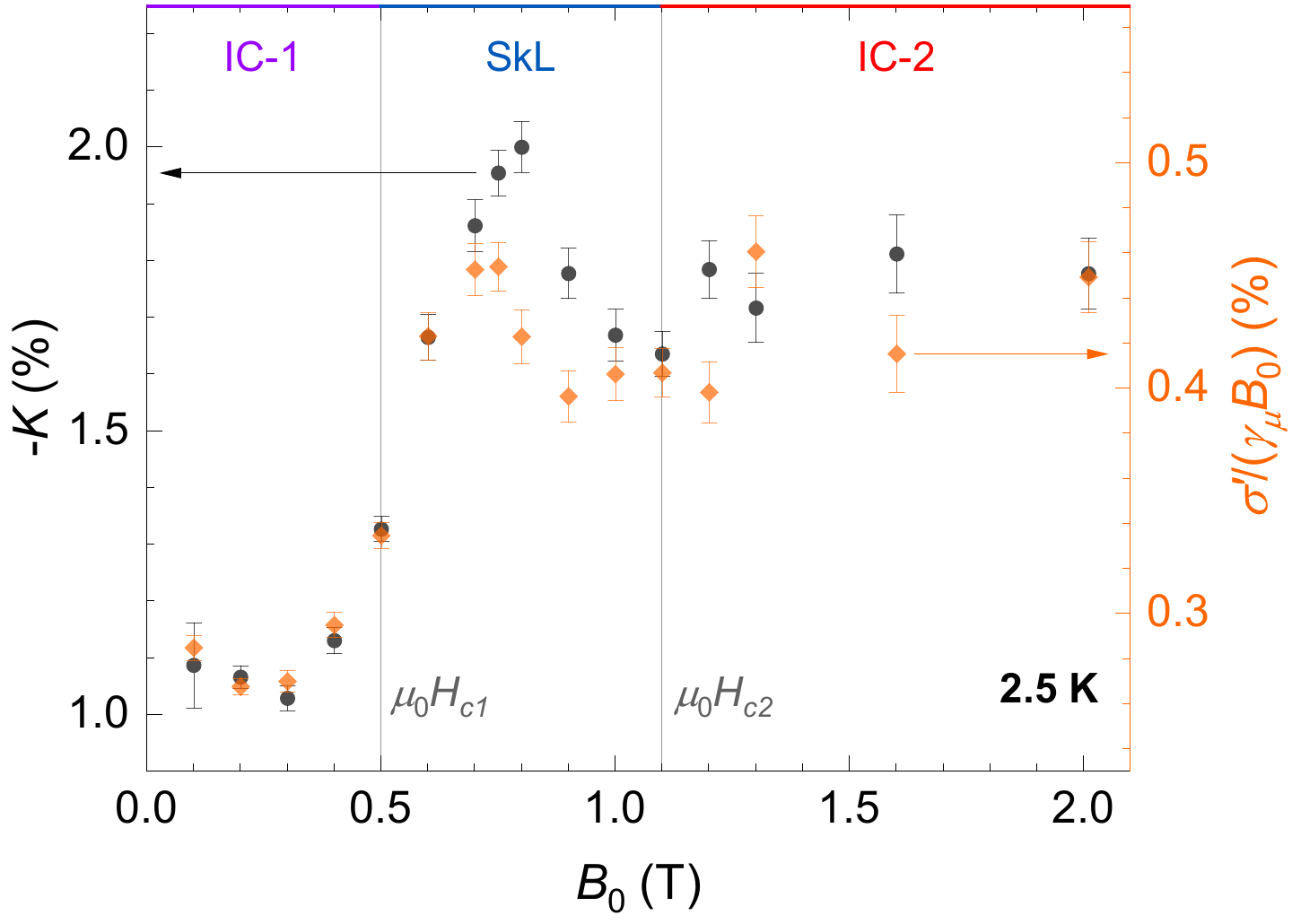}
\caption{%
Field dependence of the muon \knight shift (black), and the distribution width of \quasistatic local fields at the muon site normalized by the applied field (orange), at $T = \SI{2.5}{\kelvin}$ under \tf. 
Vertical lines highlight the critical fields.%
}
\label{fig_skl_bscan}
\end{figure}
We performed field-dependent \tf~\musr measurements at a base $T = \SI{2.5}{\kelvin}$ on the \instrHAL instrument at the S$\mu$S. 
In contrast to \instrGPS measurements, the initial muon spin orientation pointed at an angle of $\theta \approx \SI{90}{\degree}$ from the $c$ axis [see Fig. 1(b) in the main text] to maximize the number of muon decays measured by detectors along the \tDir direction. 
Early-time \tDir-direction data were fitted using a model consisting of a sum of two damped oscillations, 
$A(t) = a_s \cos(\omega t + \varphi) e^{-\sigma'^2 t^2} + a_\mathrm{bgd} \cos(\omega_0 t + \varphi_0)$, 
the first due to the sample, and the second due to the silver background. 
Here, the fitted silver-background angular frequency $\omega_0$ gave an estimate of the applied magnetic field $B_0 = \mu_0 H = \omega_0 / \gamma_\mu$, whereas the fitted sample angular frequency $\omega$ yielded the average magnitude of the local magnetic field $\braket{B} = \omega / \gamma_\mu$ at the muon site (see section \Quasistatic polarization). 
From these, the local muon \knight shift was calculated as $K = (\braket{B} - B_0)/B_0 = \omega/\omega_0 - 1$~\cite{blundell2021muon}. 
Note that the sample contribution to the muon asymmetry in this model follows \cref{eq_quasistatic_model_polarization} with the usual \tf approximation $p_r \approx 0$~\cite{blundell2021muon}. 
The latter can be justified when the applied \tf field, which points along $c$, is much larger than internal fields components parallel to the detector direction $z = \tDirEq \perp c$, and hence also when $|K| \ll 1$, 
which results in $\braket{\hat{B}_c^2} \approx 1$ and thus $p_r = \braket{\hat{B}_z^2} \leq 1 - \braket{\hat{B}_c^2} \approx 0$ \via \cref{eq_quasistatic_relaxing_asymmetry}. 


The obtained muon \knight shift (which is indeed much smaller than $\SI{100}{\percent}$, justifying the model we used) is shown in \cref{fig_skl_bscan}, alongside the local-field distribution width $\sigma' / \gamma_\mu$ [\cref{eq_quasistatic_sigma}] divided by the applied field $B_0$. 
The obtained critical applied fields $\mu_0 H_{c1} = \SI{0.5(1)}{\tesla}$ and $\mu_0 H_{c2} = \SI{1.1(1)}{\tesla}$ are consistent with AC susceptibility [Fig. 1(a) in the main text]~\cite{kurumaji2019skyrmion,hirschberger2020high,hirschberger2020topological}. 
We see that in the \icOne and \icTwo phases $K \propto \sigma'/B_0 \approx \const$, which implies $B - B_0 \propto \sigma' \propto B_0$ (\ie, the local average magnetization is proportional to the width of the local-magnetic-field distribution, which is also approximately proportional to the applied magnetic field). 
On the other hand, $K \not\propto \sigma'/B_0 \neq \const$ in the \skl phase, with both quantities showing clear peaks for applied fields near the middle of the phase. 
A similar peak in \quasistatic magnetism is also observed in other \skl hosts near the middle of their \skl phases~\cite{franke2018magnetic} (\eg, in \noncentrosymmetric \gavs and \gavse as a function of $T$). 
However, intriguingly, the peaks of $K$ and $\sigma'/B_0$ in \gdpdsi appear to be slightly shifted in field relative to each other, with the one in $K$ appearing at $\SI{0.70(5)}{\tesla}$ and the one in $\sigma'/B_0$ appearing at $\SI{0.80(5)}{\tesla}$. 

\section{\Abinitio muon site calculations}

\begin{figure*}[!tb] 
\centering
\includegraphics[width=0.95\linewidth]{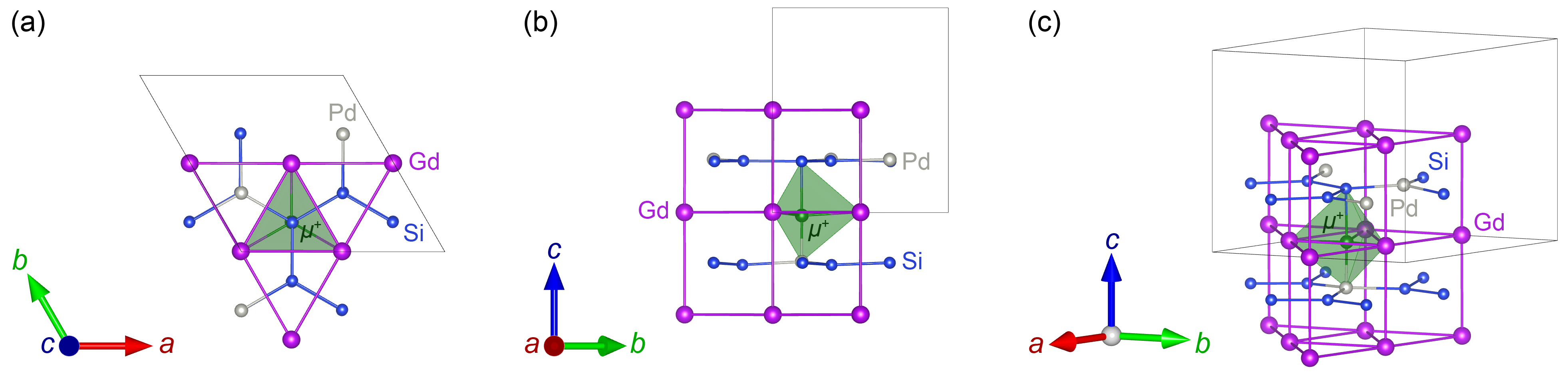}
\caption{%
Lowest-energy muon (\muonPlus) stopping site in \gdpdsi (green). 
Shown are two neighbouring hexagonal \pdAtom/\siAtom (gray/blue, respectively) superstructure layers C and D around a triangular $ab$ plane of \gdIon ions (purple) in which the muon stops.%
}
\label{sfig_muon_sites}
\end{figure*}
Muon stopping sites were determined using the \mufinder program~\cite{huddart2022mufinder}, which was used to run, cluster, and analyse the results of \abinitio muon-stopping-site calculations performed using the \castep \dftLongFirst software~\cite{clark2005first}. 
For these, we used the \pbe \xcFuncLong~\cite{perdew1996generalized} and ultrasoft pseudopotentials. 
\gdpdsi forms a \centrosymmetric superstructure with \gdpdsiFullStacking stacking of interlayer \pdAtom and \siAtom atomic configurations~\cite{tang2011crystallographic,kurumaji2019skyrmion}. 
Muon sites were calculated in a reduced superstructure with just two \pdAtom/\siAtom layers in the two most common layer stackings: \gdpdsiAB and \gdpdsiCD, as these two stackings already account for \SI{50}{\percent} of pairwise \pdAtom/\siAtom layer combinations in the full superstructure. 
The reduced unit cell dimensions of $8.16 \times 8.16 \times \SI{8.19}{\angstrom^3}$ were still large enough to avoid finite-size effects around the implanted muon. 
A $\SI{490}{\electronvolt}$ \planewaveAdj energy cutoff and \twoCubed \monkpack grid~\cite{monkhorst1976special} reciprocal-space sampling was chosen to achieve \dft convergence. 
Total energies were converged to within $\SI{e-5}{\electronvolt/\mathrm{atom}}$ in the \scfLongFirst \dft loop, while muon-site geometry was converged to within a maximum force tolerance of $\SI{0.05}{\electronvolt/\angstrom}$ on both the muon and the nuclei. 

In all of the considered local stacking configurations, muons were found to occupy the same symmetric position at the centre of an $ab$-plane \gdIon triangle (\cref{sfig_muon_sites}), with only a small displacement of $-0.29$ and ${+}\SI{0.14}{\angstrom}$ along the out-of-plane ($c$-axis) direction under \gdpdsiAB and \gdpdsiCD \pdAtom/\siAtom layer stacking, respectively. 
Due to the similarity between the stackings, we expect that all \gdIon layers in \gdpdsi host a muon site of this kind. 

\section{Magnetic fields at the muon site}

\begin{figure}[!tb] 
\centering
\includegraphics[width=0.9\columnwidth]{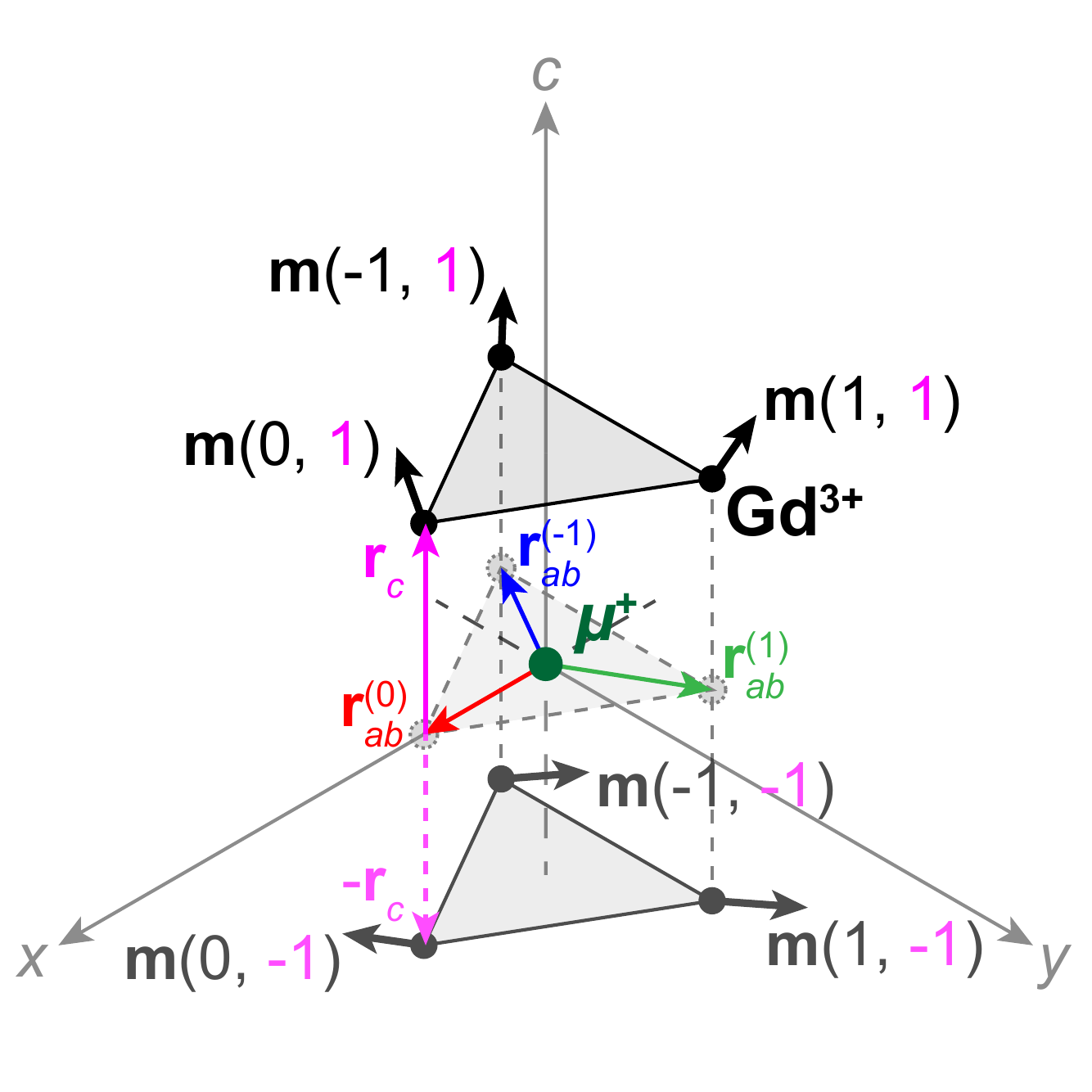}
\caption{%
Geometry of the muon site from \cref{sfig_muon_sites} showing the magnetic moments $\myvec{m}(j, {\pm}1)$ of the six \gdIon spins discussed in the text. 
Here, the local $x$ axis is defined as pointing along the $\myvec{r}_{ab}^{(0)}$ direction.%
}
\label{sfig_muon_site_geometry}
\end{figure}
\begin{figure*}[!tb] 
\centering
\includegraphics[width=1.00\linewidth]{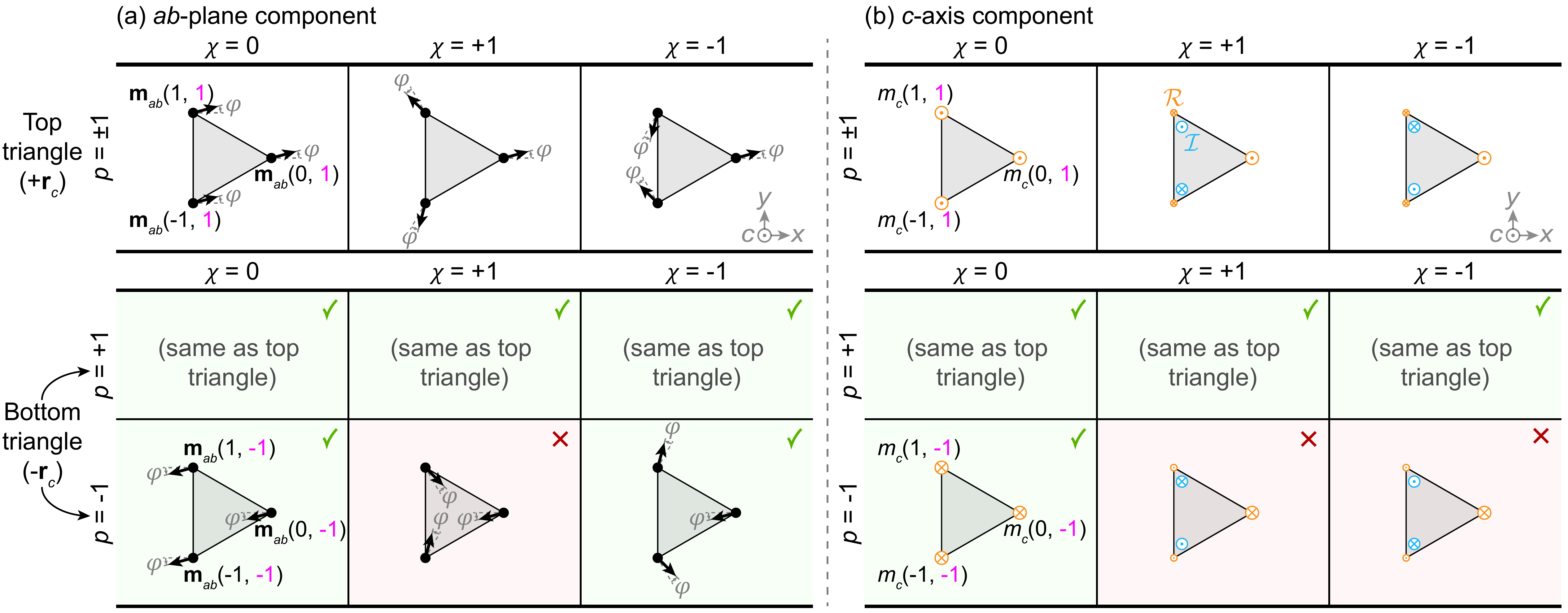}
\caption{%
Symmetry-adapted decomposition [\cref{eq_decomp_chiral}] of (a) $ab$-plane and (b) $c$-axis components of magnetic moments in a sextet of \gdIon spins comprised of two triangles (top and bottom) shown in \cref{sfig_muon_site_geometry}. 
At a given reflection parity $p$ and chirality $\chi$, the amplitude of a contribution is given by: (a) a real vector $\myvec{\tilde{m}}_{ab}(\chi, p) = \myvec{m}_{ab}(0, 1)$ in the $ab$ plane, oriented at an angle $\varphi$ from the $\myvec{r}_{ab}^{(0)}$ ($x$-axis) direction, or (b) a complex number $\tilde{m}_c(\chi, p) = m_c(0, 1)$, here shown if real and positive. 
Diameters of circles on panel (b) are proportional to the real ($\mathcal{R}$, orange) and imaginary ($\mathcal{I}$, cyan) parts of complex $c$-axis magnetic moment contributions $m_c(j, \pm 1)$, while their positive or negative sign is indicated by a dot or a cross, respectively, inside the circle. 
Cell shading indicates whether the total dipolar contribution to the magnetic field at the muon site for a given $p$ and $\chi$ is direction-preserving (green shading and check mark) or direction-mixing (red shading and cross).
As argued below, direction-mixing contributions to the are strongly suppressed in \gdpdsi.%
}
\label{sfig_muon_site_filtering}
\end{figure*}
Magnetic fields at the muon site in \gdpdsi originate from \gdIon spins. 
In this section, we will show that, to a good approximation, in-plane ($ab$-plane) and out-of-plane ($c$-axis) components of these spins generate exclusively in-plane ($\myvec{B} \perp c$) and out-of-plane ($\myvec{B} \parallel c$) local magnetic fields at the muon site, respectively. 
We will call contributions to the magnetic field at the muon site that obey this property direction-preserving, while we will call those that do not direction-mixing. 
We will show that direction-mixing contributions are almost negligible in \gdpdsi due to the nature of its muon site and its magnetic state.

Firstly, effective magnetic fields from isotropic hyperfine contact interactions of the muon's spin with the local electron spin density (which are usually small) are all direction-preserving~\cite{blundell2021muon}. 
Secondly, 
dipolar fields from \gdIon spins in the same $ab$ plane as the muon are also all direction-preserving. 
This includes the three dominant, nearest-neighbouring (NN) \gdIon spins found at a distance $r_1 = \SI{2.35}{\angstrom}$ from the muon, as well as the three next-nearest-neighbouring \gdIon spins.
Thus, the only potentially-direction-mixing magnetic fields could come from \gdIon spins outside of the muon's local $ab$ plane, with the closest being 
next-next-nearest-neighbouring (NNNN) \gdIon spins, found at a distance $r_3 = \SI{4.83}{\angstrom}$ from the muon.
However, these can contribute at most $(r_1/r_3)^3 = \SI{12}{\percent}$ as much to the magnetic field at the muon site as the dominant, NN \gdIon spins, just from the decrease of dipolar field strength with distance. 

Moreover, the local muon-site symmetry further strongly suppresses any direction-mixing contributions from \gdIon spins outside of the muon's local $ab$ plane. 
Namely, ignoring atoms other than \gdAtom, 
the muon site lies on both a mirror plane $ab$ and an axis of \threefold symmetry $c$ (\cref{sfig_muon_sites}), which causes most potentially-direction-mixing contributions to cancel at the muon position. 
To see this, we will consider a general sextet of \gdIon spins with magnetic moments $\myvec{m}(j, {\pm}1)$ displaced from the muon by $\myvec{r}(j, {\pm}1) = \myvec{r}_{ab}^{(j)} \pm \myvec{r}_c$, where $\myvec{r}_{ab}^{(j)} = R^j \myvec{r}_{ab}^{(0)}$ is the displacement within the $ab$ plane, $R$ is a rotation about the $+c$ axis by $\SI{120}{\degree}$, $j = -1, 0, 1$, and $\pm \myvec{r}_c$ is the displacement along the $c$ axis (see \cref{sfig_muon_site_geometry}). 
Note that due to \gdpdsi's crystal structure, every \gdIon ion can be assigned to a unique sextet relative to the muon, and that all \gdIon spins in a sextet are 
at equal distance 
from the muon. 
We can uniquely decompose the $ab$-plane and $c$-axis components of magnetic moments in a \gdIon sextet 
as
\begin{equation}
\begin{aligned}
\myvec{m}_{ab}(j, \pm 1) &= \sum_{p=1,-1} p^{(1 \mp 1)/2} \sum_{\chi=-1}^{1} R^{j \chi} \myvec{\tilde{m}}_{ab}(\chi, p) , \\
m_c(j, \pm 1)            &= \sum_{p=1,-1} p^{(1 \mp 1)/2} \sum_{\chi=-1}^{1} \omega^{j \chi} \tilde{m}_c(\chi, p) . 
\end{aligned}
\label{eq_decomp_chiral}
\end{equation}
where $\omega = e^{2 \pi i / 3}$ is a third root of unity, and $\myvec{\tilde{m}}_{ab}(\chi, p)$ and $\tilde{m}_c(\chi, p)$ are $ab$-plane and $c$-axis amplitudes, respectively, of contributions with parity $p = 1, -1$ (symmetric and antisymmetric, respectively) under reflections over the muon's $ab$ plane, and chirality $\chi = -1, 0, 1$ (negative, achiral, and positive, respectively) under \threefold rotations around the muon's local $c$ axis. 
Here, amplitudes $\myvec{\tilde{m}}_{ab}(\chi, p)$ are arbitrary real vectors in the $ab$ plane, 
achiral $\tilde{m}_c(0, p)$ are arbitrary real numbers, and
positive-chirality $\tilde{m}_c(1, p)$ are arbitrary complex numbers. 
On the other hand, negative chirality amplitudes $\tilde{m}_c(-1, p) = \tilde{m}_c(1, p)^*$ are fully determined \via complex conjugation $^*$ from positive chirality amplitudes $\tilde{m}_c(1, p)$, since the magnetic moments $\myvec{m}(j, \pm 1)$ produced by \cref{eq_decomp_chiral} must be real vectors. 
The patterns of magnetic moments in a \gdIon spin sextet in individual $p$ and $\chi$ contributions are shown in \cref{sfig_muon_site_filtering}. 

To see which $p$ and $\chi$ contributions are direction-mixing, we write the dipolar magnetic field at the muon site due to a magnetic moment $\myvec{m}$ displaced from the muon by $\myvec{r}$ as 
\begin{equation}
\myvec{B} = \frac{\mu_0}{4 \pi r^3} \left( \frac{3}{r^2} [ \myvec{r} \otimes \myvec{r} ] \myvec{m} - \myvec{m} \right) \propto \frac{1}{r^3} ,
\label{eq_dipolar1}
\end{equation}
where $\mu_0$ is the vacuum permeability, $r = |\myvec{r}|$, $\myvec{a} \otimes \myvec{b} = \myvec{a} \myvec{b}\tran$ denotes the outer product of column vectors $\myvec{a}$ and $\myvec{b}$, and $\tran$ denotes a vector transpose. 
Here, only the first term can be direction-mixing. 
For a \gdIon magnetic moment at position $\myvec{r} = \myvec{r}(j, {\pm}1) = \myvec{r}_{ab}^{(j)} \pm \myvec{r}_c$, we can expand it as 
\begin{equation}
\myvec{r} \otimes \myvec{r} = [ \myvec{r}_{ab}^{(j)} \otimes \myvec{r}_{ab}^{(j)} + \myvec{r}_c \otimes \myvec{r}_c ] \pm [ \myvec{r}_{ab}^{(j)} \otimes \myvec{r}_c + \myvec{r}_c \otimes \myvec{r}_{ab}^{(j)} ] .
\label{eq_dipolar2}
\end{equation}
Here, the terms in the first square bracket are direction-preserving, while those in the second square bracket 
are direction-mixing.
Summing up the dipolar fields at the muon site from all \gdIon spins in a sextet, we find that only antisymmetric ($p = -1$) contributions can be direction-mixing, and of those only positive-chirality ($\chi = +1$) $ab$-plane magnetic moments and chiral ($\chi = \pm 1$) $c$-axis magnetic moments are actually direction mixing. 
Explicitly, we find that the $c$-axis and $ab$-plane components of the magnetic field at the muon site due to direction-mixing are proportional to 
$|\myvec{\tilde{m}}_{ab}(1, -1)| \cos \varphi$ and 
$|\tilde{m}_c(1, -1)|$, respectively, where $\varphi$ is the angle between vectors $\myvec{\tilde{m}}_{ab}(1, -1)$ and $\myvec{r}_{ab}^{(0)}$ [see \cref{sfig_muon_site_filtering}(a)]. 

We next argue that the amplitudes of these direction-mixing contributions for \gdIon spins close to the muon site are actually negligible in \gdpdsi. 
Firstly, previous studies found predominantly ferromagnetic correlations between NN \gdIon layers in \gdpdsi~\cite{bouaziz2022fermi,paddison2022magnetic}, implying that $\myvec{m}(j, 1) \approx \myvec{m}(j, -1)$ for layers close to the muon, and thus that 
symmetric ($p = 1$) contributions, which are direction-preserving, dominate. 
%
We note that, as \gdIon spins in a sextet are always an even number of \gdAtom layers apart, even the presence of antiferromagnetic NN \gdAtom layer correlations would lead to the same conclusion. In other words, 
only intermediate-\wavelength inter-layer ($c$-axis) correlations could produce a direction-mixing magnetic field at the muon site. 
Secondly, since the $ab$-plane modulation \wavelengths of \incommensurate spin structures in \gdpdsi are $\lambda \approx \SI{25}{\angstrom}$~\cite{kurumaji2019skyrmion,li2023magnetic}, and thus much larger than the nearest-neighbour \gdIon-ion distance of $d_{\gdgd} = \SI{4.08}{\angstrom} \ll \lambda$~\cite{tang2011crystallographic,kurumaji2019skyrmion}, we expect that achiral contributions ($\chi = 0$; \ie, locally approximately-ferromagnetic), which are direction-preserving, dominate. 
%
The amplitudes of direction-mixing contributions, which are both antisymmetric ($p = -1$) and chiral ($\chi = \pm 1$) at the same time, are thus expected to be strongly suppressed in \gdpdsi.

In summary: (i) direction-mixing contributions to the local magnetic field at the muon site can only come from far-away \gdIon spins (NNNN or further, and in a different $ab$ plane than the muon), with their dipolar fields suppressed with distance $r$ as $(r_1/r)^3 \ll 1$, 
and (ii) the amplitudes of direction-mixing contributions 
in the local magnetic structure around the muon 
are strongly suppressed by the relatively-long-wavelength modulation of \gdIon magnetic moments in \gdpdsi (twice separately, both due to long wavelengths along the $c$-axis and within the $ab$-plane). 
We can thus safely neglect any direction-mixing contributions to the local magnetic field at the muon site in the analysis of \musr data presented in the main text. 

%